\documentclass[twocolumn,amsmath,amssymb,aps,prd,nofootinbib]{revtex4}

\usepackage{graphicx}
\usepackage{color} 
\usepackage{soul}  

\begin{document}

\title{Triple unification of inflation, dark energy, and dark matter
\\
in two-scalar-field cosmology}

\author{Paulo M. S\'a}

\email{pmsa@ualg.pt}

\affiliation{Departamento de F\'{\i}sica, Faculdade de Ci\^encias e
Tecnologia, Universidade do Algarve, Campus de Gambelas, 8005-139 Faro,
Portugal}

\begin{abstract}
A unified description of inflation, dark energy, and dark matter is presented
within a two-scalar-field cosmological model. Inflation, assumed to be of the
warm type, is driven by one of the scalar fields, which, shortly after the end
of the inflationary period, decouples from radiation and begins to oscillate
rapidly around the minimum of its potential, thus behaving like cold dark
matter; the second scalar field emerges, at recent times, as the dominant
component of the universe, giving rise to a second era of accelerated
expansion. For certain values of the parameters of the model, the cosmological
solutions arising in this triple unification of inflation, dark energy, and
dark matter are viable, reproducing the main features of the evolution of the
universe.
\end{abstract}

\maketitle

\section{Introduction\label{introduction}}

The theory of cosmic inflation
\cite{starobinsky-1980,guth-1981,linde-1982,albrecht-1982,linde-1983}
stands now on solid observational foundations, as several of its key predictions
have been confirmed by precise measurements of the
cosmic microwave background radiation \cite{Planck-inflation-2015}.

According to the inflationary paradigm, the universe undergoes a period of
accelerated expansion in the early stages of its evolution,
which is driven by a scalar field --- the inflaton --- slowly rolling down its potential.
Such an early inflationary period not only solves the flatness, horizon, homogeneity,
isotropy, and primordial monopole problems,
but also provides the seeds for the formation of the observed large-scale structures
of the universe.

A period of accelerated expansion is not exclusive to the early stages of
evolution of the universe. In fact, cosmological observations have shown, not
without surprise, that accelerated expansion is also taking place at the
present time \cite{riess-1998,perlmutter-1999}, implying the existence of an
unknown form of energy
--- dubbed as dark energy --- which accounts for a substantial part of the total
energy density of the universe \cite{Planck-parameters-2015}.
Within the $\Lambda$CDM concordance model, this dark energy is assumed to be
a cosmological constant.
However, this simple explanation raises problems of its own \cite{weinberg-1989},
a circumstance that led to the hypothesis that dark energy could be identified
with a scalar field \cite{caldwell-1998}, as in the inflationary paradigm.

In addition to dark energy, the concordance cosmological model also includes
cold dark matter, which accounts for about one quarter of the total energy density
of the universe \cite{Planck-parameters-2015}.
Although the existence of dark matter has been inferred by its gravitational
effects on a multiplicity of astrophysical and cosmological phenomena,
it has so far eluded a direct detection and, after decades of intense experimental
efforts, its physical nature remains a mystery \cite{bertone-2018}.
Such circumstances led to the consideration of a wider range of dark-matter candidates,
including the possibility that a scalar field, similar to those appearing in the models
of inflation and dark energy, could play the role of dark matter.

Since inflation, dark energy, and dark matter can all be identified with
scalar fields, it is natural to try to unify these seemingly disparate phenomena
under the same theoretical roof using these fields.

Such a unified description was proposed in Refs.~\cite{liddle-2006,liddle-2008}.
There, inflation was assumed to be of the usual (cold) type,
followed by a post-inflationary reheating period,
in which the decay of the inflaton field was required to be incomplete,
leaving a remnant that behaved like cold dark matter.
In order to reduce the energy density of the remnant to the level required by
cosmological observations, two possibilities were considered:
modification of the decay rate during the reheating process \cite{liddle-2006} or
introduction of an additional period of thermal inflation,
driven by a separate field, at lower energy densities \cite{liddle-2008}.
In what concerns dark energy, this scenario assumed a non-zero vacuum energy
for the inflaton/dark-matter field, motivated by a combination of the string
landscape picture and the anthropic principle.

A new scenario for a triple unification, in which dark energy is described not
by a cosmological constant, but rather by a dynamical scalar field, was soon
afterwards proposed \cite{henriques-2009}. Within a two-scalar-field
cosmological model inspired by supergravity, one of the fields played the role
of dark energy, inducing the present accelerated expansion of the universe,
while the second field played the roles of both inflaton and dark matter.
Because inflation was assumed to be of the warm type \cite{berera-1995}, no
distinctive post-inflationary reheating phase was required; soon after the
smooth transition to the radiation-dominated era, the energy transfer from the
inflaton field to the radiation bath ceased and the former began to oscillate
around the minimum of the potential, thus mimicking the behavior of a
cold-dark-matter fluid. However, despite its success in unifying inflation,
dark energy, and dark matter within a single framework, this two-scalar-field
cosmological model was not entirely satisfactory, since it accounted for just
a fraction of the dark matter content of the universe.

The purpose of the present article is to provide a unified description
of inflation, dark energy, and dark matter in a more general setting, namely,
within a two-scalar-field cosmological model
given by the action\footnote{Throughout this article we will adopt the natural
system of units and use the notation
$\kappa\equiv \sqrt{8\pi G}= \sqrt{8\pi}/m_\texttt{P}$,
where $G$ is the gravitational constant and
$m_\texttt{P} = 1.22 \times 10^{19}\, {\rm GeV}$ is the Planck mass.}
\begin{align}
 S = {} & \int d^4x \sqrt{-g} \bigg[
  \frac{R}{2\kappa^2} - \frac12 (\nabla \phi)^2
    \nonumber
  \\
  & - \frac12 e^{-\alpha\kappa\phi} (\nabla \xi)^2
  - e^{-\beta\kappa\phi} V(\xi) \bigg],
 \label{action 2SF}
\end{align}
where $g$ is the determinant of the metric $g_{\mu\nu}$, $R$ is the Ricci
scalar, $\phi$ and $\xi$ are scalar fields, and $\alpha$ and $\beta$ are
independent dimensionless parameters\footnote{The triple unifications of
inflation, dark energy, and dark matter proposed in
Refs.~\cite{liddle-2006,liddle-2008} and Ref.~\cite{henriques-2009} correspond
to models given by action~(\ref{action 2SF}) with, respectively,
$\alpha=\beta=0$, $\phi=0$, and $V(\xi)=V_0+\frac12 M^2 \xi^2$, where $V_0$
and $M$ are arbitrary constants, and $\alpha=0$, $\beta=-\sqrt2$, and
$V(\xi)=A_1[1-2A_2 \exp(-\sqrt2\kappa \xi)+A_3 \exp(-2\sqrt2\kappa \xi)]$,
where the constants $A_i$ are related to fundamental quantities of the
Salam-Sezgin six-dimensional supergravity theory.}. Such an action, with a
non-standard kinetic term and an exponential potential, arises in a great
variety of gravity theories, such as the Jordan-Brans-Dicke theory,
Kaluza-Klein theories, $f(R)$-gravity, and string theories (see
Refs.~\cite{berkin-1991,starobinsky-2001} for a derivation of the above action
in the context of these theories). More recently, it has been shown that this
action also arises in the context of hybrid metric-Palatini theories of
gravity \cite{harko-2012,tamanini-2013}.

For an appropriate choice of the potential $V(\xi)$, the two-scalar-field
cosmological model given by action~(\ref{action 2SF}) allows for a unified
description of inflation, dark energy, and dark matter, in which the scalar
field $\xi$ plays the roles of both inflaton and dark matter, while the scalar
field $\phi$ plays the role of dark energy.

The simplest potential providing such a triple unification has the form
\begin{align}
 V(\xi) & = V_a + \frac12 m^2 \xi^2,
  \label{potential xi}
\end{align}
where $V_a$ and $m$ are constants, related, respectively, to the energy density
of dark energy and to the $\phi$-dependent mass of the scalar field
$\xi$, defined as
\begin{align}
 M_\xi^2 (\phi) & = m^2 e^{-\beta\kappa\phi}.
   \label{mass xi}
\end{align}
While emphasizing that a triple unification such as the one proposed in this
article could be achieved by any potential whose expansion around its minimum
has the form $A+B\xi^2+\dots$, for definiteness we will use the potential
given by Eq.~(\ref{potential xi}).

In what follows, let us briefly outline the key aspects of the triple unification
proposed in this article.

Inflation is assumed to be of the warm type.
Energy is continuously transferred from the inflaton field $\xi$
(and also from the dark-energy field $\phi$) to a radiation bath,
thereby ensuring that the energy density of the latter is substantial
--- albeit sub-dominant --- throughout the inflationary expansion
and that a smooth transition to a radiation-dominated era takes place
without the need for a distinctive post-inflationary reheating phase.

The dissipation coefficients, mediating the energy trans\-fer from the scalar
fields to the radiation bath, have a generic dependence on the temperature,
namely, $\Gamma \propto T^p$ ($p$ constant), and, immediately after the end of
the inflationary period, are exponentially suppressed, becoming negligible
shortly afterwards.

Shortly after the end of the inflationary period, the inflaton $\xi$ decouples from
radiation and begins to oscillate rapidly around the minimum of its potential,
thus behaving on average like a pressureless nonrelativistic fluid, i.e.,
like cold dark matter.

Due to the non-standard kinetic term and the exponential factor in the potential,
the energy density of cold dark matter depends explicitly on the scalar field $\phi$,
implying that, in general, this quantity does not evolve exactly
as ordinary baryonic matter.

After a radiation-dominated era, encompassing the primordial nucleosynthesis period,
cold dark matter, together with ordinary baryonic matter,
dominates the dynamics of the universe, giving rise to a matter-dominated era,
long enough to allow for structure formation.

At recent times, the scalar field $\phi$ finally emerges as the dominant component
of the universe, giving rise to a second era of accelerated expansion,
thus behaving like dark energy.

These key aspects of the proposed triple unification will be detailed
in the body of the article.

To conclude this introductory section, let us point out that unified
descriptions of inflation, dark energy, and dark matter have been proposed in
several other
contexts~\cite{capozziello-2006,bose-2009,santiago-2011,odintsov-2019,lima-2019,ketov-2020,odintsov-2020}.

This article is organized as follows. The evolution equations for the
two-scalar-field cosmological model are presented in the next section. For
clarity, the cosmic evolution is divided into two stages, the first
corresponding to the inflationary period and the transition to the
radiation-dominated era (Sect.~\ref{1st stage}) and the second encompassing
the radiation-, matter-, and dark-energy-dominated eras (Sect.~\ref{2nd
stage}). The continuity of the different physical quantities at the transition
between the first and second stages of evolution is analyzed in
Sect.~\ref{transition}. Numerical solutions are presented in
Sect.~\ref{numerical}, which is divided into three subsections. In the first,
we analyze the case $\alpha=\beta$, while the second is devoted to the case
$\alpha\neq\beta$. Dissipative effects during inflation are analyzed in the
last subsection. Finally, in Sect.~\ref{conclusions}, we present our
conclusions.

\section{Two-scalar-field cosmological model\label{2SF model}}

Our analysis of the cosmic evolution is divided into two stages: the first
corresponds to the inflationary period and the transition to the
radiation-dominated era, while the second encompasses the radiation-, matter-,
and dark-energy-dominated eras.

\subsection{First stage of evolution: the inflationary era\label{1st stage}}

We assume inflation to be of the warm type (for reviews, see
Refs.~\cite{berera-2009,bastero-gil-2009}).
In this inflationary paradigm, a continuous transfer of energy from the
inflaton field to radiation ensures that the energy density of the latter
remains substantial --- albeit sub-dominant --- throughout the inflationary era.
This energy transfer also guarantees that the
transition to a radiation-dominated era takes place in a smooth manner.
It contrasts with the usual (cold) inflationary paradigm,
in which radiation is severely diluted during inflation, a circumstance that
makes a distinctive post-inflationary reheating process necessary in order
to recover the standard cosmic evolution.

In our model, radiation is described by a perfect fluid with an equation-of-state
parameter $w_\texttt{R}=p_\texttt{R}/\rho_\texttt{R}=1/3$, where $p_\texttt{R}$
and $\rho_\texttt{R}$ are the pressure and the energy density of the fluid, respectively.

The energy density of radiation is sustained, during the inflationary period,
by a continuous transfer of energy from the scalar fields $\xi$ and $\phi$,
which is  accomplished by the introduction of dissipative terms with
coefficients $\Gamma_\xi$ and $\Gamma_\phi$ into the equations of motion. This
energy transfer prevents the radiation bath from being diluted, keeping the
expanding universe ``warm''.

To implement warm inflation, it would suffice to have a continuous and
significative transfer of energy from the inflaton field $\xi$ to the
radiation bath, in which case one could simply set the dissipation coefficient
$\Gamma_\phi$ to zero and ignore any energy exchange between the dark-energy
field $\phi$ and radiation. However, as it follows from action (\ref{action
2SF}), a direct transfer of energy between the two scalar fields also takes
place (for nonvanishing $\alpha$ and/or $\beta$). Therefore, it seems natural
to also allow for a direct energy transfer from the scalar field $\phi$ to the
radiation bath, mediated by a non-zero $\Gamma_\phi$, although we shall
emphasize that this is not essential for the implementation of the
warm-inflation scenario.

Inflation comes to an end when, due to an increase of dissipative effects,
the energy density of the radiation bath smoothly takes over and begins to
dominate the evolution of the universe.
At this point, the dissipation coefficients $\Gamma_\xi$ and $\Gamma_\phi$ are
exponentially suppressed and, consequently, the radiation bath decouples from
the scalar fields $\xi$ and $\phi$ and begins to evolve in the usual manner.

Let us now present the equations governing the cosmic evolution during the
inflationary period and the transition to a radiation-dominated era.

We assume a flat Friedman-Robertson-Walker universe\footnote{Since the
current cosmological measurements constrain the present-time value of the
curvature density parameter $\Omega_k$ to be very small
\cite{Planck-parameters-2015}, a spatially flat universe can be assumed without
much loss of generality.}, given by the metric
\begin{align}
ds^2 = -dt^2 + a^2(t) d\Sigma^2,
  \label{FRW metric}
\end{align}
where $a(t)$ is the scale factor and $d\Sigma^2$ is the metric of the
three-dimensional Euclidean space.

The equations of motion for the scalar fields $\xi(t)$ and $\phi(t)$ and for
the energy density of radiation $\rho_\texttt{R}(t)$ are then
\begin{align}
 & \ddot{\xi} + 3 \frac{\dot{a}}{a} \dot{\xi} - \alpha\kappa \dot{\phi}\dot{\xi}
 + \frac{\partial V}{\partial \xi} e^{(\alpha-\beta)\kappa\phi}
 = - \Gamma_\xi \dot{\xi} e^{\alpha\kappa\phi}, \label{ddot-xi-1}
  \\
 & \ddot{\phi} + 3 \frac{\dot{a}}{a} \dot{\phi} + \frac{\alpha\kappa}{2}
 \dot{\xi}^2 e^{-\alpha\kappa\phi}
 -\beta \kappa V e^{-\beta \kappa \phi}
 = - \Gamma_\phi \dot{\phi}, \label{ddot-phi-1}
  \\
 & \dot{\rho_\texttt{R}} + 4 \frac{\dot{a}}{a} \rho_\texttt{R} = \Gamma_\xi
 \dot{\xi}^2 + \Gamma_\phi \dot{\phi}^2, \label{dot-rho-1}
\end{align}
while the Einstein equations for the scale factor $a(t)$ are given by
\begin{align}
& \bigg( \frac{\dot{a}}{a} \bigg)^2 = \frac{\kappa^2}{3} \bigg(
 \frac{\dot{\phi}^2}{2} + \frac{\dot{\xi}^2}{2} e^{-\alpha\kappa\phi}
 + V e^{-\beta\kappa\phi} + \rho_\texttt{R} \bigg), \label{dot-a-1}
   \\
& \hspace{2.7mm}\frac{\ddot{a}}{a} =  - \frac{\kappa^2}{3} \bigg( \dot{\phi}^2 +
 \dot{\xi}^2 e^{-\alpha\kappa\phi} - V e^{-\beta\kappa\phi} + \rho_\texttt{R} \bigg),
 \label{ddot-a-1}
\end{align}
where an overdot denotes a derivative with respect to time $t$ and the
potential $V$ is given by Eq.~(\ref{potential xi}).

Note that the above evolution equations differ from the usual ones in warm
inflationary models in that they contain extra terms arising due to the presence,
in action (\ref{action 2SF}), of a non-standard kinetic term for the field $\xi$.

Instead of the comoving time $t$, let us use a new variable $u$, related to the
redshift $z$,
\begin{align}
  u = -\ln \left( \frac{a_0}{a} \right) = - \ln (1+z),
  \label{variable u}
\end{align}
where $a_0\equiv a(u_0)$ denotes the value of the scale factor at the
present time $u_0=0$.

With this change of variables, the above equations for $\xi$, $\phi$, and
$\rho_\texttt{R}$ become
\begin{align}
  &  \hspace{-0.5mm}  \xi_{uu} = - \bigg\{
     \bigg[ \frac{\ddot{a}}{a} + 2 \bigg( \frac{\dot{a}}{a} \bigg)^2
           +\frac{\dot{a}}{a} \Gamma_\xi e^{\alpha\kappa \phi} \bigg]\xi_u
    \nonumber
  \\
  & \hspace{8.5mm}
    - \alpha\kappa  \bigg( \frac{\dot{a}}{a} \bigg)^2 \phi_u \xi_u
    + m^2 \xi e^{(\alpha-\beta)\kappa \phi}
  \bigg\} \bigg( \frac{\dot{a}}{a} \bigg)^{-2},
      \label{Eq xi E1}
  \\
  & \hspace{-1.0mm}\phi_{uu} =  - \bigg\{
     \bigg[ \frac{\ddot{a}}{a} + 2 \bigg( \frac{\dot{a}}{a} \bigg)^2
           +\frac{\dot{a}}{a} \Gamma_\phi
     \bigg]\phi_u
  + \frac{\alpha\kappa}{2} \bigg( \frac{\dot{a}}{a} \bigg)^2 \xi_u^2
  e^{-\alpha\kappa\phi} \nonumber
  \\
  & \hspace{8.5mm}
   -\beta\kappa \left( V_a +\frac12 m^2 \xi^2 \right)
   e^{-\beta\kappa\phi}
  \bigg\} \bigg( \frac{\dot{a}}{a} \bigg)^{-2},
      \label{Eq phi E1}
  \\
  & \hspace{0mm} \rho_{\texttt{R}u} = - 4 \rho_\texttt{R}
  + \frac{\dot{a}}{a} \left( \Gamma_\xi \xi_u^2 + \Gamma_\phi \phi_u^2 \right),
      \label{Eq rho E1}
\end{align}
where the subscript $u$ denotes a derivative with respect to $u$; $\dot{a}/a$
and $\ddot{a}/a$ are functions of $u$, $\xi$, $\xi_u$, $\phi$, and $\phi_u$,
given by
\begin{align}
 \left( \frac{\dot{a}}{a} \right)^2 = 2\kappa^2
 \frac{ \left( V_a + \frac12 m^2 \xi^2 \right) e^{-\beta\kappa\phi} +
 \rho_\texttt{R} }{6 - \kappa^2 \phi_u^2
 - \kappa^2 \xi_u^2 e^{-\alpha\kappa\phi}}
 \label{Eq friedman E1}
\end{align}
and
\begin{align}
& \frac{\ddot{a}}{a} = \frac{\kappa^2}{3}
 \Bigg\{ 2\kappa^2 \left[ \left( V_a+\frac12 m^2 \xi^2 \right)
 e^{-\beta\kappa\phi} + \rho_\texttt{R} \right] \nonumber
 \\
& \hspace{6.4mm} \times \frac{\phi_u^2 + \xi_u^2 e^{-\alpha\kappa\phi}}
 {\kappa^2 \phi_u^2 + \kappa^2 \xi_u^2 e^{-\alpha\kappa\phi} - 6}
 \nonumber
 \\
& \hspace{6.4mm} + \left( V_a + \frac12 m^2 \xi^2 \right)
  e^{-\beta\kappa\phi} - \rho_\texttt{R}
    \Bigg\}.
   \label{Eq Dotdota E1}
\end{align}

In order to solve the above system of equations, one has to specify the
dissipation coefficients $\Gamma_\xi$ and $\Gamma_\phi$.

Over the years, a variety of forms has been adopted for these coefficients,
from the simplest, based on general phenomenological considerations, to the
more elaborate ones, derived from microscopic quantum field theory. In
general, the dissipation coefficients $\Gamma$ appearing in the literature are
functions of the temperature $T$ and/or the inflaton field $\xi$, as, for
instance, $\Gamma \propto T^3/\xi^2$
\cite{bastero-gil-2013,lima-2019}, $\Gamma \propto T$
\cite{bastero-gil-2016,rosa-2019,rosa-2019b}, or $\Gamma \propto T^{-1}$
\cite{bastero-gil-2019}.

In this article, we will not be concerned with the specific microscopic models
used to derive the dissipation coefficients. We will adopt instead a
model-independent approach, assuming that, during inflation, these
coefficients have a generic dependence on the temperature of the radiation
bath, namely, $\Gamma \propto T^p$. We also assume that, immediately after the
end of the inflationary period, the dissipation coefficients are exponentially
suppressed, becoming negligible soon afterwards. In short, we assume
the dissipation coefficients $\Gamma_\xi$ and $\Gamma_\phi$ to be given by
\begin{align}
 \Gamma_{\xi,\phi} = f_{\xi,\phi} \times \left\{
    \begin{aligned}
     & T^p,
     & T\geq T_\texttt{E},
    \\
     & T^p  \exp \left[ 1- \left( \frac{T_\texttt{E}}{T} \right)^q \right],
     & T\leq T_\texttt{E},
    \end{aligned}
 \right.
 \label{gammas}
\end{align}
where $T_\texttt{E}$ is the temperature of the radiation bath at the end of
the inflationary period, $f_\xi$ and $f_\phi$ are positive constants with
dimension $(\textrm{mass})^{1-p}$ encoding the details of the microscopic
models used to derive the dissipation coefficients, and $q>0$ and $p$ are
parameters determining the temperature dependence of these coefficients.

The suppression of the dissipation coefficients immediately after the
inflationary period is of paramount importance in our unification proposal in
order to guarantee that the inflaton field $\xi$ survives and that it has
enough energy to mimic the behavior of cold dark matter in a way consistent
with cosmological observations (see Sect.~\ref{2nd stage} below for details).
A microscopic model in which this suppression is achieved naturally has been
proposed recently \cite{rosa-2019}, in the context of the warm little inflaton
scenario \cite{bastero-gil-2016}. More specifically, the dissipation
coefficient mediating the energy transfer from the inflaton field to
radiation, which initially is proportional to $T$, becomes exponentially
suppressed when the temperature drops below a certain threshold value (roughly
coinciding with the end of the inflationary period), thus leading to a stable
inflaton remnant. Such suppression of the dissipative effects below a
threshold temperature is also present in another warm-inflation model
\cite{bastero-gil-2019} in which the dissipation coefficient during inflation
is proportional to $T^{-1}$.

As will be shown in Sect.~\ref{num-dissipative},
for dissipation coefficients proportional to $T^p$, with $p>2$,
suppression of the dissipation coefficients below a threshold
temperature occurs naturally, as a result of the background dynamics,
making it unnecessary, for such values of p, to introduce explicitly the
exponential factor in Eq.~(\ref{gammas}).

For our base scenario we will choose $p=1$,
corresponding to dissipation coefficients linearly dependent on the temperature,
and $q=2$ (see Sect.~\ref{num-base}); other values of the parameters $p$ and $q$
will be considered in Sect.~\ref{num-dissipative}.

Finally, let us recall that the temperature $T$ of the radiation bath is
related to its energy density by
\begin{align}
 \rho_\texttt{R} = \frac{\pi^2}{30}g_* T^4,
   \label{temp-rho}
\end{align}
where $g_*$ denotes the effective number of relativistic degrees of freedom at
temperature $T$. Assuming the standard model of particle physics and taking
into account that, at the relevant temperatures, all the degrees of freedom of
this model are relativistic and in thermal equilibrium, $g_*$ takes the value
$106.75$.

Solving Eqs.~(\ref{Eq xi E1})--(\ref{Eq Dotdota E1}) allows us to determine
the density parameters for radiation and for the scalar fields $\xi$ and
$\phi$,
\begin{align}
& \Omega_\texttt{R} = \frac{\rho_\texttt{R}}{\rho_c}=
  \frac{\kappa^2}{3}\rho_\texttt{R} \left( \frac{\dot{a}}{a} \right)^{-2},
\\
& \Omega_\xi = \frac{\rho_\xi}{\rho_c}= \frac{\kappa^2}{6} \bigg[
  \xi_u^2 \, e^{-\alpha\kappa\phi}
  + m^2 e^{-\beta\kappa\phi} \xi^2
  \left( \frac{\dot{a}}{a} \right)^{-2} \bigg],
\\
& \Omega_\phi = \frac{\rho_\phi}{\rho_c}= \frac{\kappa^2}{3} \bigg[
  \frac{\phi_u^2}{2} + V_a e^{-\beta\kappa\phi}
  \left( \frac{\dot{a}}{a} \right)^{-2} \bigg], \label{Omega-phi-E1}
\end{align}
as well as the effective equation-of-state parameter,
\begin{align}
 w_{\rm eff} = \frac13 \bigg( \Omega_\texttt{R}
 + 3 \Omega_\phi \frac{p_\phi}{\rho_\phi}
 + 3 \Omega_\xi \frac{p_\xi}{\rho_\xi} \bigg), \label{eos parameter E1}
\end{align}
where $\rho_c=(3/\kappa^2)(\dot{a}/a)^2$ is the critical density and the
energy density and pressure of the scalar fields $\xi$ and $\phi$ are given by,
respectively,
\begin{align}
 \rho_\xi = \frac12 \left( \frac{\dot{a}}{a} \right)^2 \xi_u^2
 e^{-\alpha\kappa\phi}
 + \frac12 m^2 e^{-\beta\kappa\phi} \xi^2,
  \label{rho xi E1}
\\
 p_\xi = \frac12 \left( \frac{\dot{a}}{a} \right)^2 \xi_u^2
 e^{-\alpha\kappa\phi}
 - \frac12 m^2 e^{-\beta\kappa\phi} \xi^2,
    \label{p xi E1}
\end{align}
and
\begin{align}
 \rho_\phi = \frac12 \left( \frac{\dot{a}}{a} \right)^2 \phi_u^2
 + V_a e^{-\beta\kappa\phi},
  \label{rho phi E1}
 \\
  p_\phi = \frac12 \left( \frac{\dot{a}}{a} \right)^2 \phi_u^2
 - V_a e^{-\beta\kappa\phi}.
  \label{p phi E1}
\end{align}

In this article we will be focused on the background dynamics of the
two-scalar-field cosmological model and its capability to reproduce the
main stages of evolution of the universe, leaving the analysis of the
primordial spectrum of density perturbations and its agreement with
cosmological observations to future work.
Nevertheless, we would like to emphasize here that warm-inflation models
have been shown to be consistent with cosmic microwave background (CMB) data
for a large variety of potentials and dissipation coefficients.
In particular, predictions of the tensor-to-scalar ratio and the spectral tilt
of the primordial spectrum were shown to agree with cosmological data for models
with a quartic potential and dissipation coefficients proportional to $T$ and $T^3$ \cite{benetti-2017,arya-2018,bastero-gil-2018,bastero-gil-2018a,motaharfar-2019},
and also in the case of a quadratic potential and dissipation
coefficient proportional to $T^{-1}$ \cite{bastero-gil-2019}.
The former models favor mostly the weak dissipative regime, while the latter is
consistent with strong dissipation.

We now turn to the description of the second stage of evolution, which
encompasses the radiation-, matter-, and dark-energy-dominated eras.

\subsection{Second stage of evolution: the radiation-, matter-, and
dark-energy-dominated eras\label{2nd stage}}

As mentioned above, at the end of the inflationary period, the dissipation
coefficients $\Gamma_\xi$ and $\Gamma_\phi$ are exponentially suppressed and,
soon afterwards, become negligible, allowing us to set them exactly to zero.
This marks the end of the first stage of evolution.

During the second stage of evolution, in the absence of dissipation, radiation
decouples from the scalar fields $\xi$ and $\phi$ and Eq.~(\ref{Eq rho E1})
yields the solution
\begin{align}
 \rho_\texttt{R}=\rho_{\texttt{R}0}\,e^{-4u},
   \label{Eq rho E2}
\end{align}
where $\rho_{\texttt{R}0}\equiv \rho_{\texttt{R}}(u_0)$ denotes the energy
density of radiation at the present time $u_0=0$.

For its part, the scalar field $\xi$ begins to oscillate rapidly around its
minimum, behaving like a nonrelativistic dark-matter fluid with equation of state
$\langle p_\xi \rangle=0$ \cite{turner-1983},
where the brackets $\langle ... \rangle$
denote the average over an oscillation\footnote{These oscillations take place
if the mass of the scalar field $\xi$, given by Eq.~(\ref{mass xi}),
is much bigger than the Hubble parameter, $M_\xi \gg H \equiv \dot{a}/a$,
a condition that can be easily satisfied by choosing a large
enough value for the constant $m$.}.

Let us derive an expression for the energy density of the dark-matter fluid
in terms of $u$ and $\phi(u)$.
To that end, we multiply Eq.~(\ref{Eq xi E1}) by $\xi_u$ and use the
definition of $\rho_\xi$ given by Eq.~(\ref{rho xi E1}) to obtain
\begin{align}
& \rho_{\xi u}
 +3 \left( \frac{\dot{a}}{a} \right)^2 \xi_u^2 e^{-\alpha\kappa\phi}
 -\frac{\alpha\kappa}{2} \left( \frac{\dot{a}}{a} \right)^2
   \xi_u^2 \phi_u e^{-\alpha\kappa\phi} \nonumber
 \\
& \hspace{15mm} + \frac{\beta\kappa}{2} m^2 \xi^2 \phi_u
  e^{-\beta\kappa\phi}
 = 0.
\end{align}

Averaging over an oscillation period and taking into account that
$\big<p_\xi\big>=0$ implies $\big< \xi^2 \big>
=\rho_\xi m^{-2} e^{\beta\kappa\phi}$ and
$\big< \xi_u^2 \big> = \rho_\xi (\dot{a}/a)^{-2} e^{\alpha\kappa\phi}$,
the above equation can be written as
\begin{align}
 \rho_{\xi u} + 3 \rho_\xi
 - \frac{(\alpha-\beta)\kappa}{2} \rho_\xi \phi_u = 0,
\end{align}
yielding the solution
\begin{align}
 \rho_\xi = C e^{-3u} e^{\frac{(\alpha-\beta)\kappa}{2}\phi},
   \label{rho dark matter}
\end{align}
where $C$ is a constant whose value is fixed by current cosmological
measurements [see Eq.~(\ref{condition 2}) below].

As expected, the energy density of dark matter is proportional to $e^{-3u}$
(or, in terms of the scale factor, proportional to $a^{-3}$), due to the fact that
the potential $V(\xi)$ was chosen to be quadratic.
But it also depends directly on the scalar field $\phi$,
through an exponential factor, as a consequence of both the non-standard
kinetic term of the scalar field $\xi$ and the exponential potential
[see action~(\ref{action 2SF})].
As will be seen in Sect.~\ref{numerical}, such dependence of $\rho_\xi$ on the
dark-energy field $\phi$ has implications on the cosmic evolution, leading
to a non-simultaneous peaking of the energy densities
of dark matter and ordinary baryonic matter.

Now, taking into account the above expressions for the energy densities of
radiation and dark matter,
the evolution Eqs.~(\ref{Eq xi E1})--(\ref{Eq Dotdota E1})
can be considerably simplified, yielding
\begin{align}
 \phi_{uu} = {}& - \bigg\{
  \bigg[ \frac{\ddot{a}}{a} + 2 \bigg( \frac{\dot{a}}{a} \bigg)^2 \bigg] \phi_u
  -\beta\kappa V_a e^{-\beta\kappa\phi} \nonumber
\\
  & +\frac{(\alpha-\beta)\kappa C}{2}
  e^{\frac{(\alpha-\beta)\kappa}{2}\phi} e^{-3u} \bigg\} \left(
\frac{\dot{a}}{a} \right)^{-2}, \label{Eq phi E2}
\end{align}
with
\begin{align}
 \left( \frac{\dot{a}}{a} \right)^2= {}&
 2\kappa^2 \bigg[ V_a e^{-\beta\kappa\phi}
 + \left( \rho_{\texttt{BM}0}
 + C e^{\frac{(\alpha-\beta)\kappa}{2}\phi} \right) e^{-3u} \nonumber
\\
 & + \rho_{\texttt{R}0}e^{-4u} \bigg]
 \left( 6-\kappa^2 \phi_u^2 \right)^{-1},
\label{Eq friedman E2}
\end{align}
and
\begin{align}
 \frac{\ddot{a}}{a} = {}& \frac{\kappa^2}{6} \bigg\{
 4\kappa^2 \bigg[ V_a e^{-\beta\kappa\phi}
     + \left( \rho_{\texttt{BM}0}
     + C e^{\frac{(\alpha-\beta)\kappa}{2}\phi} \right) e^{-3u} \nonumber
\\
 & + \rho_{\texttt{R}0} e^{-4u} \bigg]
 \phi_{u}^2 \left( \kappa^2 \phi_u^2 - 6 \right)^{-1}
 + 2 V_a e^{-\beta\kappa\phi} \nonumber
\\
 & - \left( \rho_{\texttt{BM}0}
   + C e^{\frac{(\alpha-\beta)\kappa}{2}\phi} \right) e^{-3u}
   - 2\rho_{\texttt{R}0} e^{-4u} \bigg\}.
 \label{Eq Dotdota E2}
\end{align}

In the above equations, we have introduced ordinary baryonic matter, described
as a perfect fluid with pressure $p_\texttt{BM}=0$ and energy density
\begin{align}
 \rho_\texttt{BM} = \rho_{\texttt{BM}0}\,e^{-3u},
  \label{rho baryonic matter}
\end{align}
where, as usual, the subscript $0$ indicates present-time values.

Agreement with current cosmological measurements \cite{Planck-parameters-2015}
requires $\rho_{\texttt{R}0}=9.02\times10^{-128}\, m_\texttt{P}^4$ and
$\rho_{\texttt{BM}0}=8.19\times10^{-125}\, m_\texttt{P}^4$, as well as
\begin{align}
 & \frac12 \bigg[ \left( \frac{\dot{a}}{a} \right)^2 \phi_{u}^2 \bigg]_{u=u_0}
 + V_a e^{-\beta\kappa\phi_0} = \rho_{\texttt{DE}0},
\label{condition 1}
\\
& C e^{\frac{(\alpha-\beta)\kappa}{2}\phi_0} = \rho_{\texttt{DM}0},
\label{condition 2}
\end{align}
where the present-time energy densities of dark energy and dark matter are
$\rho_{\texttt{DE}0}=1.13\times10^{-123}\, m_\texttt{P}^4$ and
$\rho_{\texttt{DM}0} = 4.25\times10^{-124}\, m_\texttt{P}^4$,
respectively\footnote{Note that these values for $\rho_{\texttt{R}0}$,
$\rho_{\texttt{BM}0}$, $\rho_{\texttt{DM}0}$, and $\rho_{\texttt{DE}0}$
correspond to a Hubble constant $H_0\equiv(\dot{a}/a)_0= 1.17 \times 10^{-61}
\, m_\texttt{P}$ or, in more familiar units, $H_0=67\,{\rm km}\,{\rm
s}^{-1}\,{\rm Mpc}^{-1}$.}.

As will be shown in Sect.~\ref{numerical}, for a choice of $V_a$ and $C$
satisfying the above conditions, Eqs.~(\ref{Eq phi E2})--(\ref{Eq Dotdota E2})
describe a radiation-dominated era, encompassing the primordial
nucleosynthesis period, followed by an era dominated by the scalar field $\xi$
(dark matter) and ordinary baryonic matter, lasting long enough for structure
formation to occur, and, finally, a dark-energy-dominated era, induced by the
scalar field $\phi$, during which the universe undergoes accelerated
expansion. The requirement that the transition from the radiation- to the
matter-dominated era does not occur too early in the cosmic history and,
consequently, does not conflict with primordial nucleosynthesis, as well as
the requirement that the expansion of the universe is accelerating at the
present time, imposes constraints on the parameters $\alpha$ and $\beta$,
namely, $|\alpha-\beta| \lesssim 1$ and $|\beta| \lesssim 3/2$.

During the second stage of evolution, the density parameter for the scalar
field $\phi$ is given by Eq.~(\ref{Omega-phi-E1}), while the density
parameters for radiation, baryonic matter, and the scalar field $\xi$, are
given by, respectively,
\begin{align}
 & \Omega_\texttt{R} = \frac{\rho_\texttt{R}}{\rho_c} =
 \frac{\kappa^2}{3}\rho_{\texttt{R}0} e^{-4u}
 \left( \frac{\dot{a}}{a} \right)^{-2},
\\
 & \Omega_\texttt{BM} = \frac{\rho_\texttt{BM}}{\rho_c} =
 \frac{\kappa^2}{3}\rho_{\texttt{BM}0} e^{-3u}
 \left( \frac{\dot{a}}{a} \right)^{-2},
\\
 & \Omega_\xi = \frac{\rho_\xi}{\rho_c} =
 \frac{\kappa^2}{3} C e^{\frac{(\alpha-\beta)\kappa}{2}\phi} e^{-3u}
 \left( \frac{\dot{a}}{a} \right)^{-2}.
\end{align}

The effective equation-of-state parameter, during this stage of evolution, is
\begin{align}
w_{\rm eff} = \frac13 \bigg[ 1 - \Omega_\texttt{BM} - \Omega_\xi - \Omega_\phi
\bigg( 1- 3 \frac{p_\phi}{\rho_\phi} \bigg) \bigg], \label{eos parameter E2}
\end{align}
where $\rho_\phi$ and $p_\phi$ are given by Eqs.~(\ref{rho phi E1}) and
(\ref{p phi E1}), respectively.

A detailed analysis of this stage of evolution for the case $\alpha=2/\sqrt6$
and arbitrary $\beta$ can be found in Ref.~\cite{sa-2020},
where a unified description of dark matter and dark energy was proposed
within the generalized hybrid metric-Palatini theory of gravity.

\subsection{Transition between the first and second stages of evolution
\label{transition}}

As we have just seen, the first stage of evolution, corresponding to the
inflationary era, is described by Eqs.~(\ref{Eq xi E1})--(\ref{Eq Dotdota
E1}), while the second stage of evolution, encompassing the radiation-, matter-,
and dark-energy-dominated eras, is described by Eqs.~(\ref{Eq phi E2})--(\ref{Eq
Dotdota E2}).

The transition from the first to the second stage of evolution occurs shortly
after the end of inflation, at the beginning of the radiation-dominated era,
when the dissipation coefficients $\Gamma_\xi$ and $\Gamma_\phi$, given by
Eq.~(\ref{gammas}), are exponentially suppressed and become negligible. The
moment at which the dissipation coefficients are set exactly to zero marks the
end of the first stage of evolution and the beginning of the second. In what
follows, this transition moment will be denoted by $u=u_*$.

At the transition between the first and second stages of evolution the
different physical quantities should be continuous.

For the scalar field $\phi$ this is achieved by
simply requiring the initial value of the second stage to be equal to the
final value of the first stage.

For the energy density of the scalar field $\xi$, continuity at the
transition requires
\begin{align}
 \rho_\xi(u_*) =
 C e^{-3u_*}
 e^{\frac{(\alpha-\beta)\kappa}{2}\phi_*},
      \label{continuity rho-xi}
\end{align}
where $\phi_*\equiv\phi(u_*)$ and $\rho_\xi(u_*)$ denotes the energy density
of the $\xi$ field during the first stage of evolution, given by Eq.~(\ref{rho
xi E1}), evaluated at $u=u_*$. Since the constant $C$ is fixed by
Eq.~(\ref{condition 2}), satisfying the above continuity condition amounts to
fix the value of $u_*$ or, equivalently, the duration of the first stage of
evolution,
\begin{align}
 \Delta^{[\texttt{I}]}u \equiv {} & u_*-u_i \nonumber
 \\
 = {} & - \frac13 \ln \frac{\rho_\xi(u_*)}{\rho_{\texttt{DM}0}}
 -\frac{(\alpha-\beta)\kappa}{6}[\phi_0-\phi_*] - u_i,
  \label{duration I}
\end{align}
where $u_i$ denotes the value of $u$ at the beginning of the first stage of
evolution.

Finally, for the energy density of radiation, continuity at the transition
requires that
\begin{align}
 \rho_\texttt{R}(u_*) = \rho_{\texttt{R}0} e^{-4u_*},
   \label{continuity rho-R}
\end{align}
where $\rho_\texttt{R} (u_*)$ denotes the energy density of radiation during
the first stage of evolution [i.e., the solution of Eq.~(\ref{Eq rho E1})]
evaluated at $u=u_*$. Because $\rho_{\texttt{R}0}$ is fixed by current
cosmological measurements (see the discussion in Sect.~\ref{2nd stage}), the
above continuity condition leads, in general, to a value of $u_*$ different
from the one determined from Eq.~(\ref{continuity rho-xi}). To avoid this and,
at the same time, to maintain adherence to the convention $u_0=0$, the value
of the variable $u$ at the beginning of the first stage of evolution, $u_i$,
has to be shifted by an appropriate amount, i.e., by an amount ensuring that
$u_*$, determined from Eq.~(\ref{continuity rho-xi}), also satisfies
Eq.~(\ref{continuity rho-R}). This procedure fixes the duration of the second
stage of evolution to be
\begin{equation}
 \Delta^{[\texttt{II}]}u \equiv u_0-u_*
 = \frac14 \ln \frac{\rho_\texttt{R}(u_*)}{\rho_{\texttt{R}0}}.
  \label{duration II}
\end{equation}

Now, we can proceed to the numerical analysis of the equations of our
two-scalar-field cosmological model.

\section{Numerical solutions\label{numerical}}

Let us solve numerically Eqs.~(\ref{Eq xi E1})--(\ref{Eq Dotdota E1}) and
Eqs.~(\ref{Eq phi E2})--(\ref{Eq Dotdota E2}), corresponding to the first and
second stages of evolution, respectively, and present a unified description of
inflation, dark energy, and dark matter within the two-scalar-field
cosmological model given by action~(\ref{action 2SF}).

For appropriate choices of the initial values for the variables $\xi$,
$\xi_u$, $\phi$, $\phi_u$, and $\rho_\texttt{R}$ and the values of the
constants $\alpha$, $\beta$, $V_a$, $m$, $C$, $f_\xi$, $f_\phi$, $p$, and $q$ our
model allows for a cosmic evolution consistent with current cosmological
observations.

Note that, because of the symmetries of action~(\ref{action 2SF}), we can assume
$\alpha\geq0$ without loss of generality. The solutions corresponding to
negative values of $\alpha$ can be obtained from the solutions with positive values
using the transformation $\alpha\rightarrow-\alpha$, $\beta\rightarrow-\beta$,
and $\phi\rightarrow-\phi$.

For clarity of presentation, this section is divided into three subsections,
where we analyze the case $\alpha=\beta$, the case $\alpha\neq\beta$, and the
dependence of the dissipative effects on the parameters $p$ and $q$.

\subsection{Case $\alpha=\beta$\label{num-base}}

We first analyze the case $\alpha=\beta=1$, choosing representative values for
the initial conditions, namely $\xi(u_i)=0.52\, m_\texttt{P}$,
$\phi(u_i)=10^{-3}\, m_\texttt{P}$, $\xi_u(u_i)=10^{-2}\, m_\texttt{P}$,
$\phi_u(u_i)=10^{-5}\, m_\texttt{P}$, $\rho_\texttt{R}(u_i)=0.25\times
10^{-12} \,m_\texttt{P}^4$, and for the parameters\footnote{The parameters
$f_\xi$ and $f_\phi$, which determine the amplitude of the energy transfer
between the scalar fields and the radiation bath, are, in general, different.
Here, for simplicity, they are assumed to be equal.}, namely,
$V_a=7.64\times10^{-123} \, m_\texttt{P}^4$, $m=10^{-5} \, m_\texttt{P}$,
$f_\xi=f_\phi=2$, $p=1$, $q=2$. We call this case the base scenario.

\begin{figure}[t]
\includegraphics[width=7.9cm]{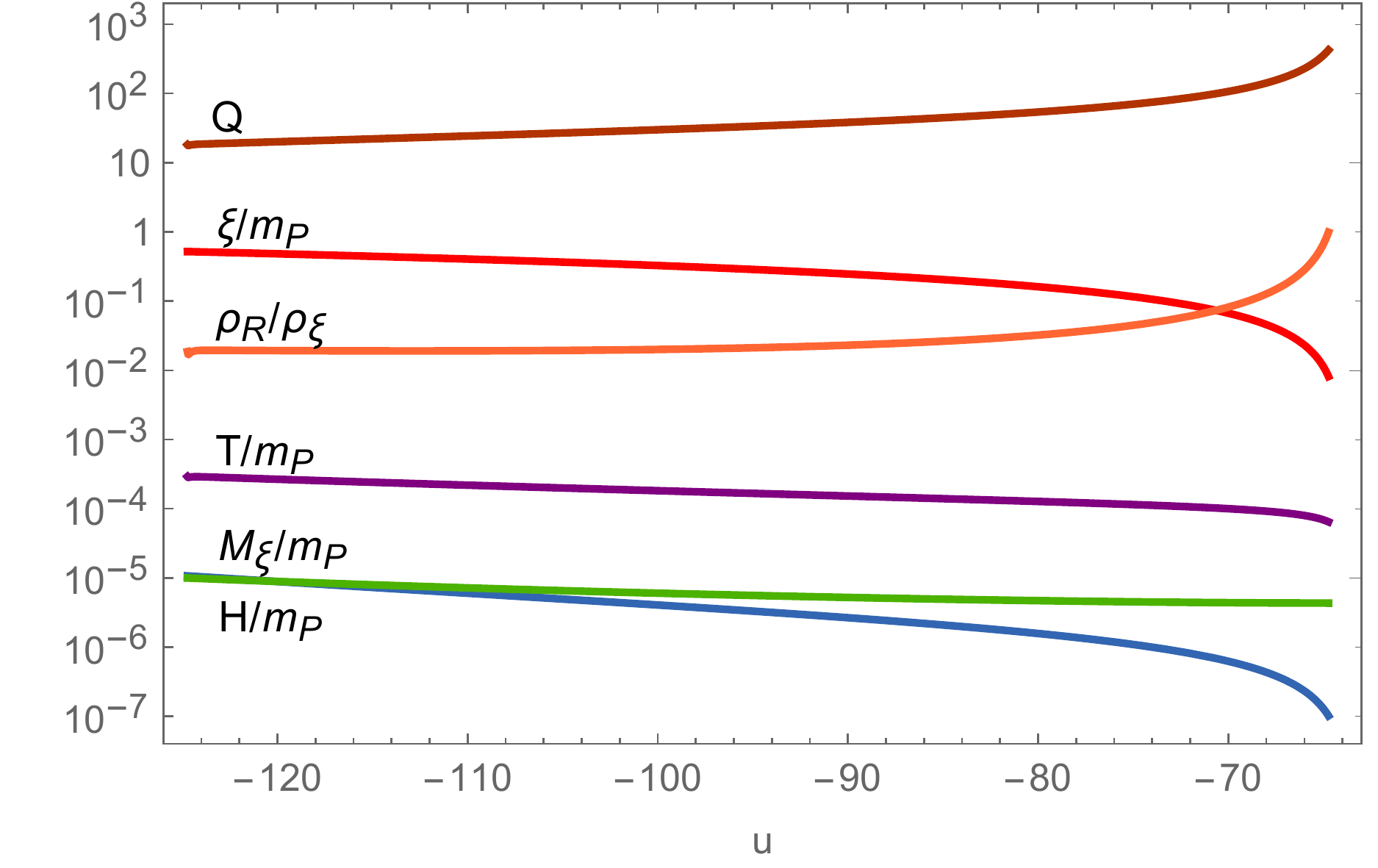}
\caption{Evolution of the inflaton field $\xi$, its mass $M_\xi$, the
dissipation ratio $Q$, the temperature $T$ of the radiation bath, the Hubble
parameter $H$, and the ratio $\rho_\texttt{R}/\rho_\xi$ during the
inflationary period, which extends from $u\approx-124.7$ to $u\approx-64.7$
($60$ $e$-folds of expansion). The energy scale of inflation is $E_{\rm
inf}\approx 2.3\times10^{16} \, {\rm GeV}$.}
 \label{Fig_inflation}
\end{figure}

Initially, the scalar field $\xi$ (the inflaton) slowly rolls down its
quadratic potential, leading to an accelerated expansion of the universe ($60$
$e$-folds). The energy scale of inflation, defined as
\begin{align}
E_{\rm inf}=\left[ e^{-\beta\kappa\phi(u_i)} V[\xi(u_i)] \right]^{1/4},
\end{align}
is $2.3\times10^{16} \, {\rm GeV}$. Because of the dependence on the scalar
field $\phi$, the inflaton mass $M_\xi$ decreases from $10^{-5} \,
m_\texttt{P}$ to $4.3\times 10^{-6} \, m_\texttt{P}$ during this stage of
cosmic evolution. Dissipative effects guarantee that energy is continuously
transferred from both fields $\xi$ and $\phi$ to the radiation bath,
preventing it from being diluted away by accelerated expansion. Throughout the
inflationary period, the dissipation ratios, defined as\footnote{Since we are
considering $\Gamma_\phi=\Gamma_\xi$, in what follows, for
simplicity, both dissipation ratios $Q_\phi$ and $Q_\xi$ will
be denoted by $Q$.}
\begin{align}
 Q_{\xi,\phi} = \frac{\Gamma_{\xi,\phi}}{3H},
   \label{q}
\end{align}
remain larger than unity (strong dissipative regime) and the temperature of
the radiation bath decreases only slightly (from $3.5\times10^{15}\, {\rm
GeV}$ to $7.9\times10^{14}\, {\rm GeV}$). The evolution of the relevant
quantities during the inflationary period is shown in
Fig.~\ref{Fig_inflation}.

\begin{figure}[t]
\includegraphics[width=7.9cm]{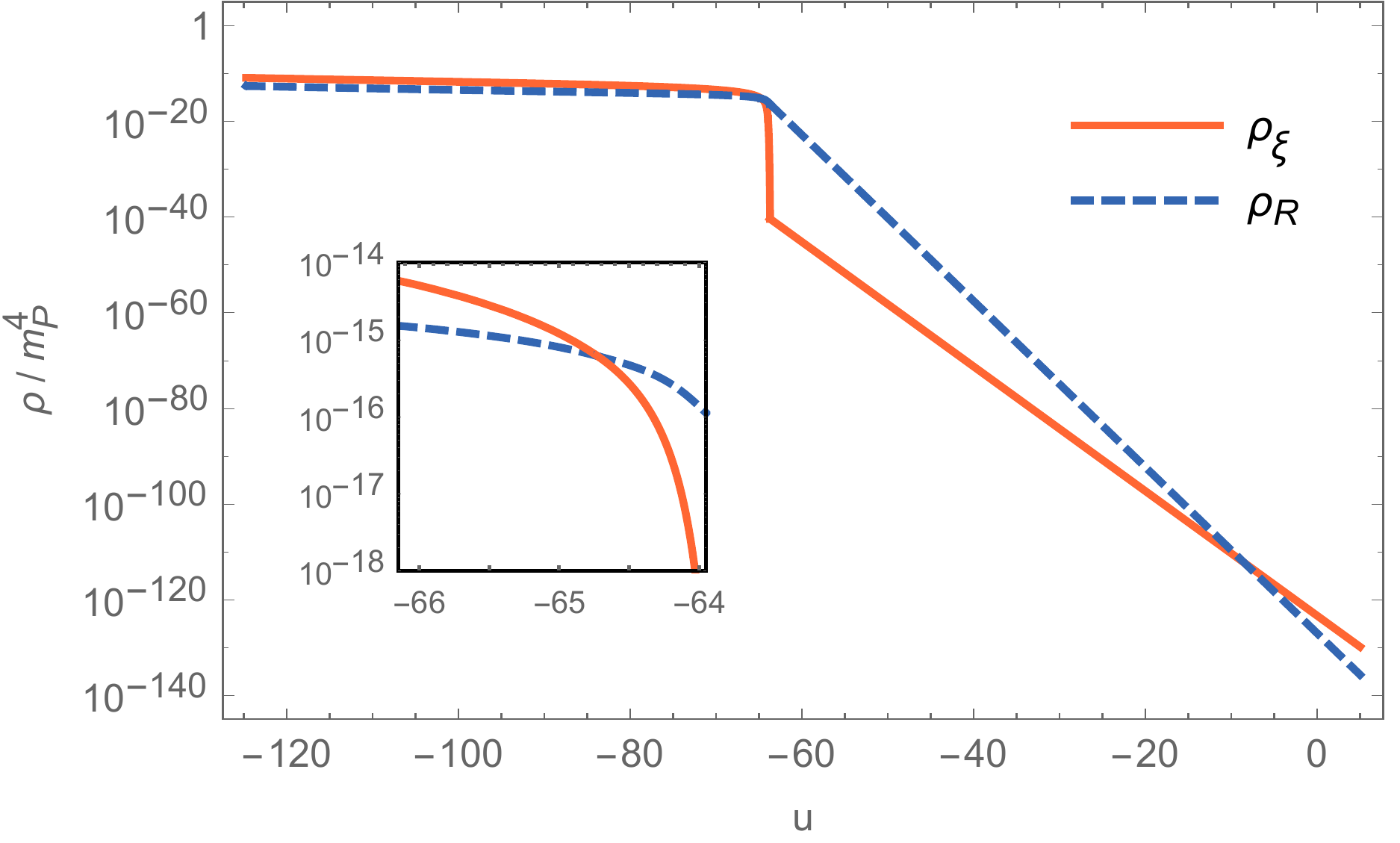}
\caption{Evolution of the energy densities of the scalar field $\xi$ and of
radiation. During the inflationary period, due to dissipative effects, the
latter remains almost constant. At $u\approx-64.7$, radiation emerges as the
dominant component of the universe and the inflationary period comes to an end
(inset plot). During the radiation-dominated era, the scalar field $\xi$
behaves like a pressureless nonrelativistic fluid (dark matter), becoming
dominant, together with ordinary baryonic matter, at $u\approx-8.6$.}
 \label{Fig_rhoxi-rhoR-E1}
\end{figure}

\begin{figure}[t]
\includegraphics[width=7.9cm]{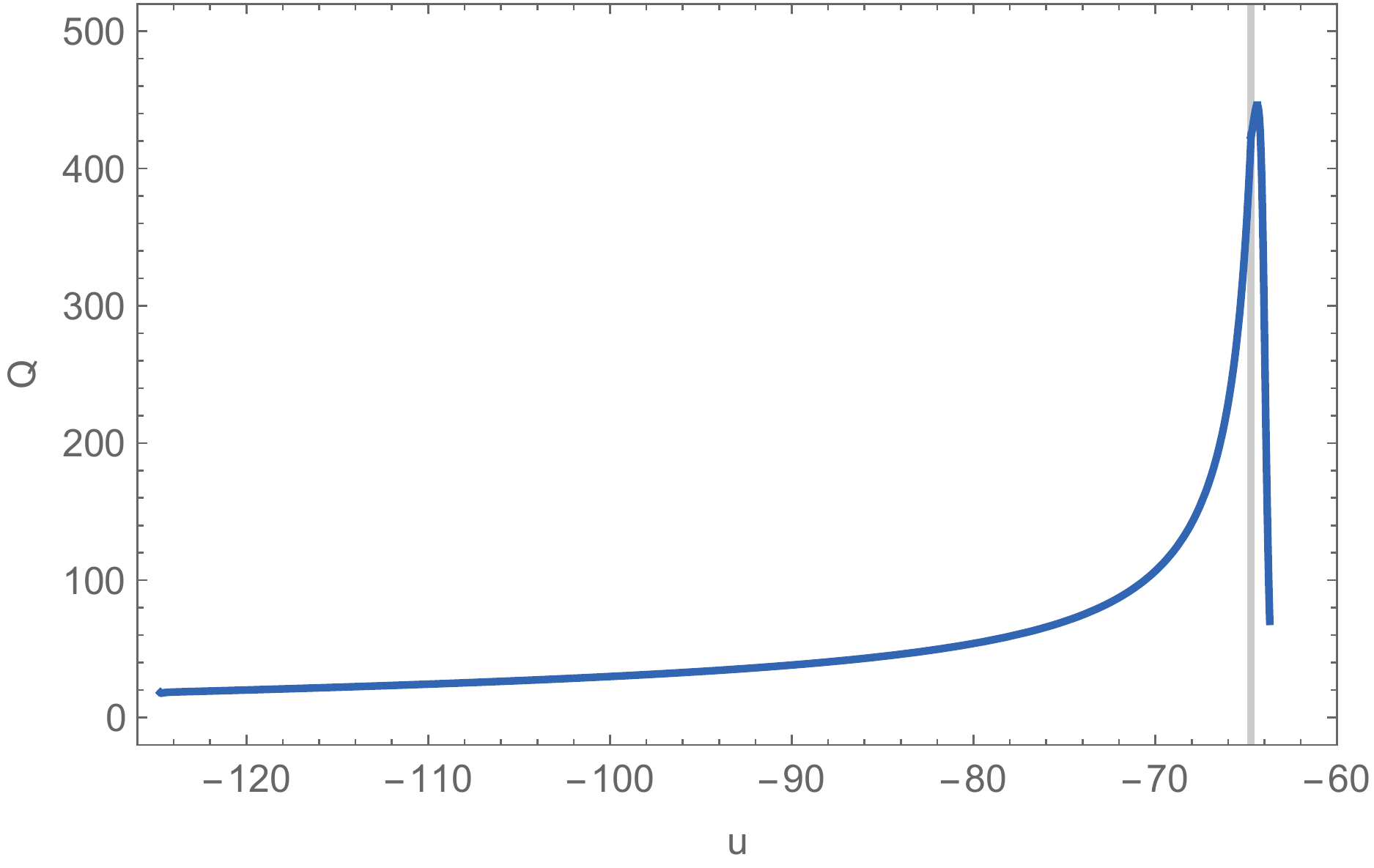}
\caption{Evolution of the dissipation ratio $Q$. Strong dissipation, $Q\gg1$,
is maintained throughout inflation. At the end of the inflationary period
($u\approx-64.7$, vertical grey line), the dissipation coefficients
$\Gamma_\xi$ and $\Gamma_\phi$ are exponentially suppressed, implying that
soon afterwards $Q$ becomes negligible. At $u=u_*\approx-63.7$, the
dissipation coefficients are set exactly to zero, marking the end of the first
stage of evolution.}
 \label{Fig-Gamma}
\end{figure}

At a certain point of the evolution, radiation emerges as the dominant
component of the universe and the inflationary period comes to an end (see
Fig.~\ref{Fig_rhoxi-rhoR-E1}). At that moment, when the temperature of the
radiation bath is $T_\texttt{E} \approx 7.9 \times 10^{14} \, {\rm GeV}$, the
dissipation coefficients $\Gamma_\xi$ and $\Gamma_\phi$ begin to decrease
exponentially, implying that soon afterwards the dissipation ratio $Q$ becomes
negligible (see Fig.~\ref{Fig-Gamma}). At $u=u_*\approx-63.7$, the dissipation
coefficients $\Gamma_\xi$ and $\Gamma_\phi$ are set exactly to zero and the
dynamics of the cosmic evolution becomes governed by Eqs.~(\ref{Eq phi
E2})--(\ref{Eq Dotdota E2}). This marks the beginning of the second stage of
evolution.

In the scenario we have been considering, the initial conditions
and the values of the parameters are such that a strong dissipative regime ($Q>1$)
is maintained throughout inflation. However, as will be shown below in
Sect.~\ref{num-dissipative},
a weak dissipative regime can also be obtained for other choices of initial
conditions and values of the parameters.

The value of the energy density of the field $\xi$ at the transition between
the first and second stages of evolution, $\rho_\xi(u_*)$, is of crucial
importance (see Fig.~\ref{Fig_rhoxi-rhoR-E1}). It cannot be too large,
otherwise the radiation-dominated era will be too short, conflicting with
primordial nucleosynthesis, but it cannot be too small, otherwise the
matter-dominated era will not be long enough for structure formation to take
place or, worse, such an era may not even occur. As discussed above [see
Eq.~(\ref{duration I})], in order to guarantee an adequate value of
$\rho_\xi(u_*)$, the duration of the first stage of evolution should be chosen
carefully; in the example we have been considering (base scenario),
$\Delta^{[\texttt{I}]}u \approx 61.0$, implying $\rho_\xi(u_*) \approx 4.2
\times 10^{-41} \, m_\texttt{P}^4$, which allows for radiation- and
matter-dominated eras with durations consistent with current cosmological
observations.

Let us open here a parenthesis to briefly comment on a recent proposal of
unification of inflation and dark matter \cite{rosa-2019}, in which inflation
is also assumed to be of the warm type. There, the potential of the inflaton
field, in addition to a quadratic term, also has a quartic one, which
dominates for large field values, implying that this scalar field behaves,
during most of the radiation-dominated era, including the primordial nucleosynthesis
period, as dark radiation. It begins to behave like dark matter
just before the transition between the radiation- and the matter-dominated eras.
This contrasts with the situation in our model,
in which the scalar field $\xi$ behaves like dark matter from the very beginning
of the radiation-dominated era [see Eq.~(\ref{rho dark matter}) and
Fig.~\ref{Fig_rhoxi-rhoR-E1}].

Returning to our base scenario, we point out that, as discussed in
Sect.~\ref{transition}, continuity of the energy density of radiation at the
transition between the first and second stages of evolution fixes the duration
of the second stage to be $\Delta^{[\texttt{II}]}u \approx 63.7$; at the
transition, $\rho_\texttt{R}(u_*) \approx 4.2 \times 10^{-17} \,
m_\texttt{P}^4$.

\begin{figure}[t]
\includegraphics[width=7.9cm]{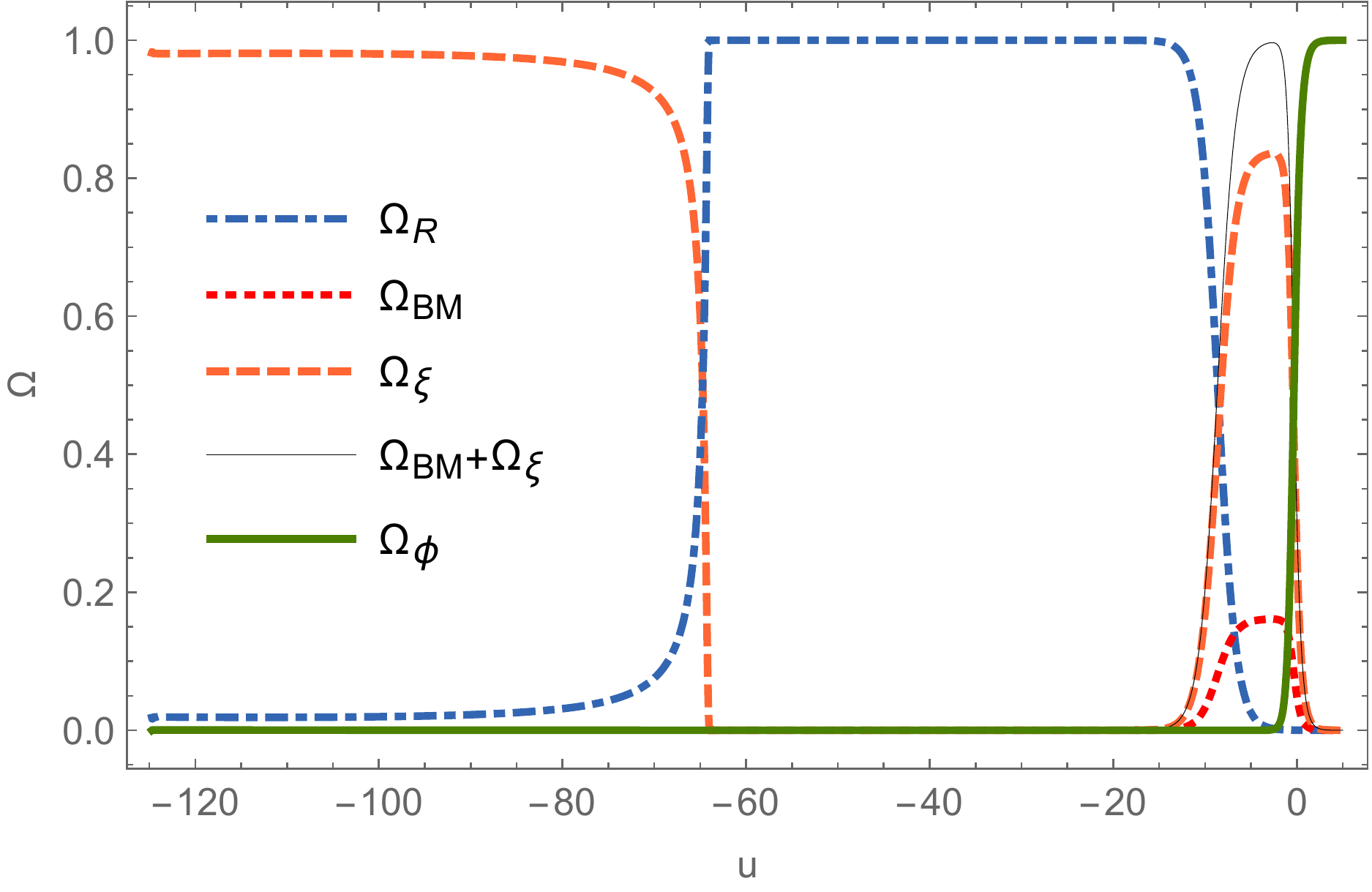}
\caption{Evolution of the density parameters for radiation, baryonic matter,
and the scalar fields $\xi$ (inflaton and dark matter) and $\phi$ (dark
energy). The successive inflationary, radiation, matter, and dark energy eras
are clearly delimited. At the present time, $u_0=0$, the density parameters
are $\Omega_{\phi}(u_0)\approx 0.69$, $\Omega_{\xi}(u_0)\approx 0.26$,
$\Omega_{\texttt{BM}}(u_0)\approx0.05$, and
$\Omega_{\texttt{R}}(u_0)\approx5.5\times10^{-5}$, in agreement with
cosmological measurements.}
 \label{Fig-Omega}
\end{figure}

During the second stage of evolution, the energy densities $\rho_\texttt{R}$ and
$\rho_\xi$ evolve according to Eqs.~(\ref{Eq rho E2}) and (\ref{rho dark matter}),
leading successively to radiation- and dark-matter-dominated eras.
Meanwhile, the scalar field $\phi$ plays no
significant role in the dynamics of the universe, only becoming dominant at
recent times ($u\approx-0.3$) when it  induces a period of accelerated expansion
(see Fig.~\ref{Fig-Omega}).

For our choice of the initial conditions and parameters (and $C \approx
4.25\times10^{-124} \, m_\texttt{P}^4$) the conditions given by
Eqs.~(\ref{condition 1}) and (\ref{condition 2}) are satisfied, implying that
the density parameters at the present time $u_0=0$ become
$\Omega_{\phi}(u_0)\approx 0.69$, $\Omega_{\xi}(u_0)\approx 0.26$,
$\Omega_{\texttt{BM}}(u_0)\approx0.05$, and
$\Omega_{\texttt{R}}(u_0)\approx5.5\times10^{-5}$, in agreement with
cosmological measurements \cite{Planck-parameters-2015}.

\begin{figure}[t]
\includegraphics[width=7.9cm]{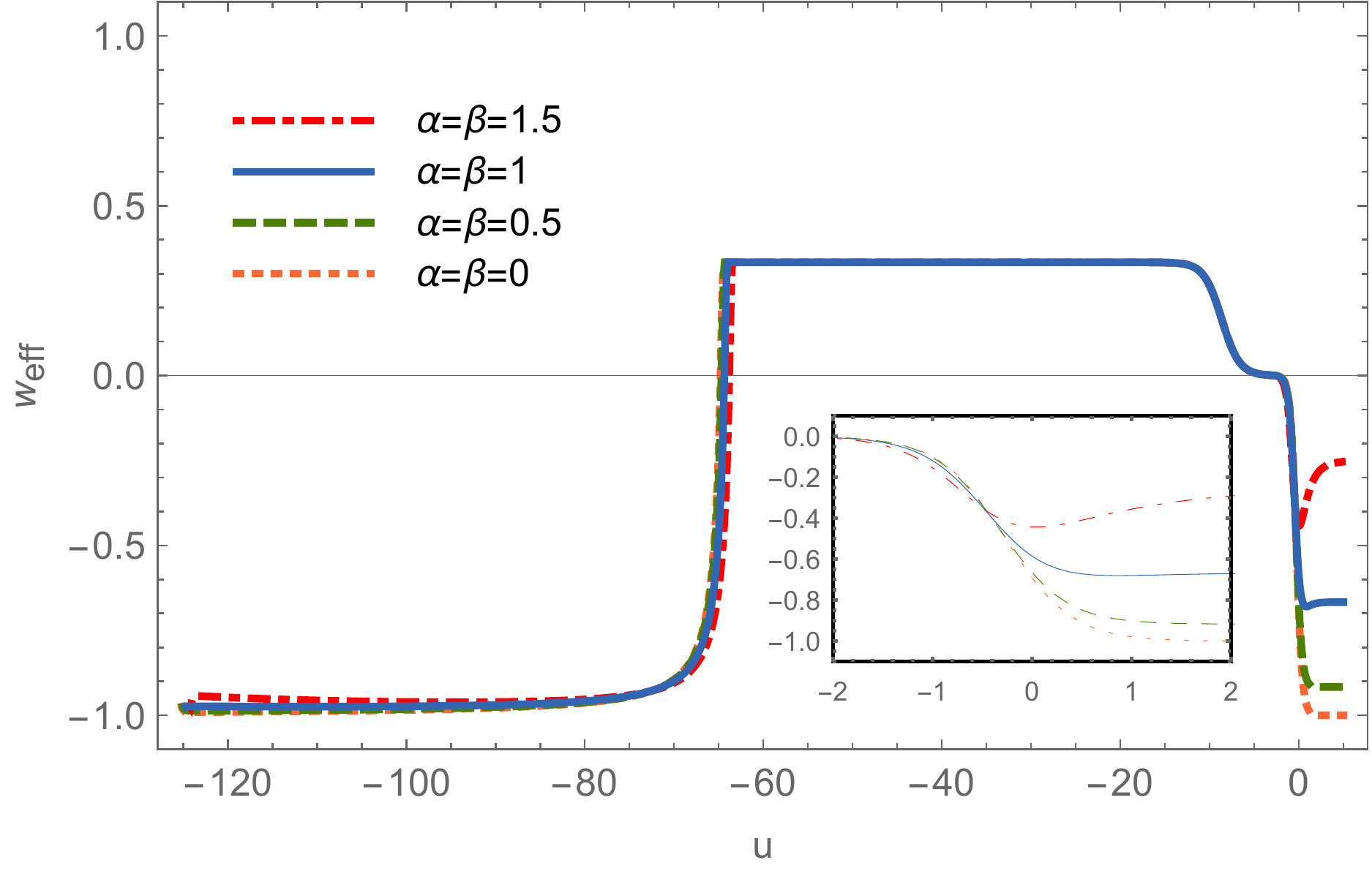}
\caption{Evolution of the effective equation-of-state parameter $w_{\rm eff}$,
clearly showing the inflationary, radiation, matter, and dark-energy eras
(the solid blue line corresponds to the base scenario, $\alpha=\beta=1$).
The value of $w_{\rm eff}$ for $u\rightarrow+\infty$ depends on the parameter
$\beta$ (but not on $\alpha$), such that, for $|\beta|<\sqrt2$, the universe
enters a period of everlasting accelerated expansion. For $\beta=3/2$ this
accelerated expansion is only temporary, since asymptotically $w_{\rm eff}$
approaches the value $-1/4$.}
 \label{Fig-weff}
\end{figure}

In the future (i.e., for $u>0$), the density parameters for radiation
$\Omega_{\texttt{R}}$, baryonic matter, $\Omega_{\texttt{BM}}$,
and dark matter $\Omega_{\xi}$ become negligible in comparison
with the density parameter for dark energy $\Omega_{\phi}$,
implying that the effective equation-of-state parameter $w_{\rm eff}$
tends to the value $-1+\beta^2/3$ (see Fig.~\ref{Fig-weff}).

Indeed, for negligible $\Omega_{\texttt{R}}$, $\Omega_{\texttt{BM}}$,
and $\Omega_{\xi}$, Eq.~(\ref{Eq phi E2}) simplifies considerably,
becoming
\begin{align}
 \phi_{uu} =  {} &- \frac{1}{2\kappa}(\kappa\phi_u -\beta) (6-\kappa^2\phi_u^2).
\end{align}
This differential equation admits an analytical solution, which, in
the limit $u\rightarrow+\infty$ and for $\beta^2<6$, becomes
$\phi_u=\beta/\kappa$.
Now, inserting this asymptotic solution, in Eq.~(\ref{Eq friedman E2}), one obtains
$V_a e^{-\beta\kappa\phi}=(6-\beta^2)H^2/(2\kappa^2)$.
Finally, substituting the above expressions for $\phi$ and $\phi_u$
into Eq.~(\ref{eos parameter E2}), one obtains
$w_{\rm eff}=-1+\beta^2/3$ for $u\rightarrow+\infty$.

The asymptotic behavior of the effective equation-of-state parameter implies that,
for $|\beta|<\sqrt2$, the universe enters a period of everlasting
accelerated expansion.
For values of $|\beta|$ slightly above $\sqrt2$, this accelerated expansion
still takes place, but does not last forever.
For instance, in the case $\beta=3/2$, shown in Fig.~\ref{Fig-weff},
accelerated expansion occurs at the present time ($w_{\rm eff}<-1/3$)
and then ceases as $w_{\rm eff}$ tends to $-1/4$.

In summary, in the example we have been considering (base scenario),
inflation, driven by the scalar field $\xi$, begins at $u\approx-124.7$ and
extends for $60$ $e$-folds, till $u\approx-64.7$. During this period, a
radiation bath with temperature of about $10^{15} \, {\rm GeV}$ is sustained
by a continuous and copious energy transfer from the scalar fields $\xi$ and
$\phi$. At the end of inflation, the dissipation coefficients are
exponentially suppressed; as a consequence, radiation decouples from the
scalar fields $\xi$ and $\phi$ and begins to evolve in the usual manner,
dominating the dynamics of the universe till $u\approx-8.6$. In the absence of
dissipative effects, the scalar field $\xi$ oscillates around its minimum,
behaving like a pressureless nonrelativistic fluid (dark matter), and,
together with ordinary baryonic matter, becomes dominant at $u\approx-8.6$.
The scalar field $\phi$ (dark energy), having played no significant role in
the dynamics of the universe during the preceding eras, finally becomes
dominant at $u\approx-0.3$, giving rise to an everlasting period of
accelerated expansion of the universe.

Because we are only interested in models that support accelerated expansion at
the present time, we restrict our analysis to the cases $\beta\lesssim3/2$.
Furthermore, as pointed out above, we can assume $\alpha\geq0$ without loss of
generality. Our numerical simulations show that, for values of $\alpha=\beta$
lying in this interval, the cosmic evolution is quite similar to the base
scenario, making it unnecessary to present here a detailed analysis. We just
refer the reader to the base scenario and also to Fig.~\ref{Fig-weff}, where,
the cosmic evolution is outlined for three more cases, namely
$\alpha=\beta=0,1/2$ and $3/2$.

\subsection{Case $\alpha\neq\beta$
\label{num-deviations}}

\begin{figure}[t]
\includegraphics[width=7.9cm]{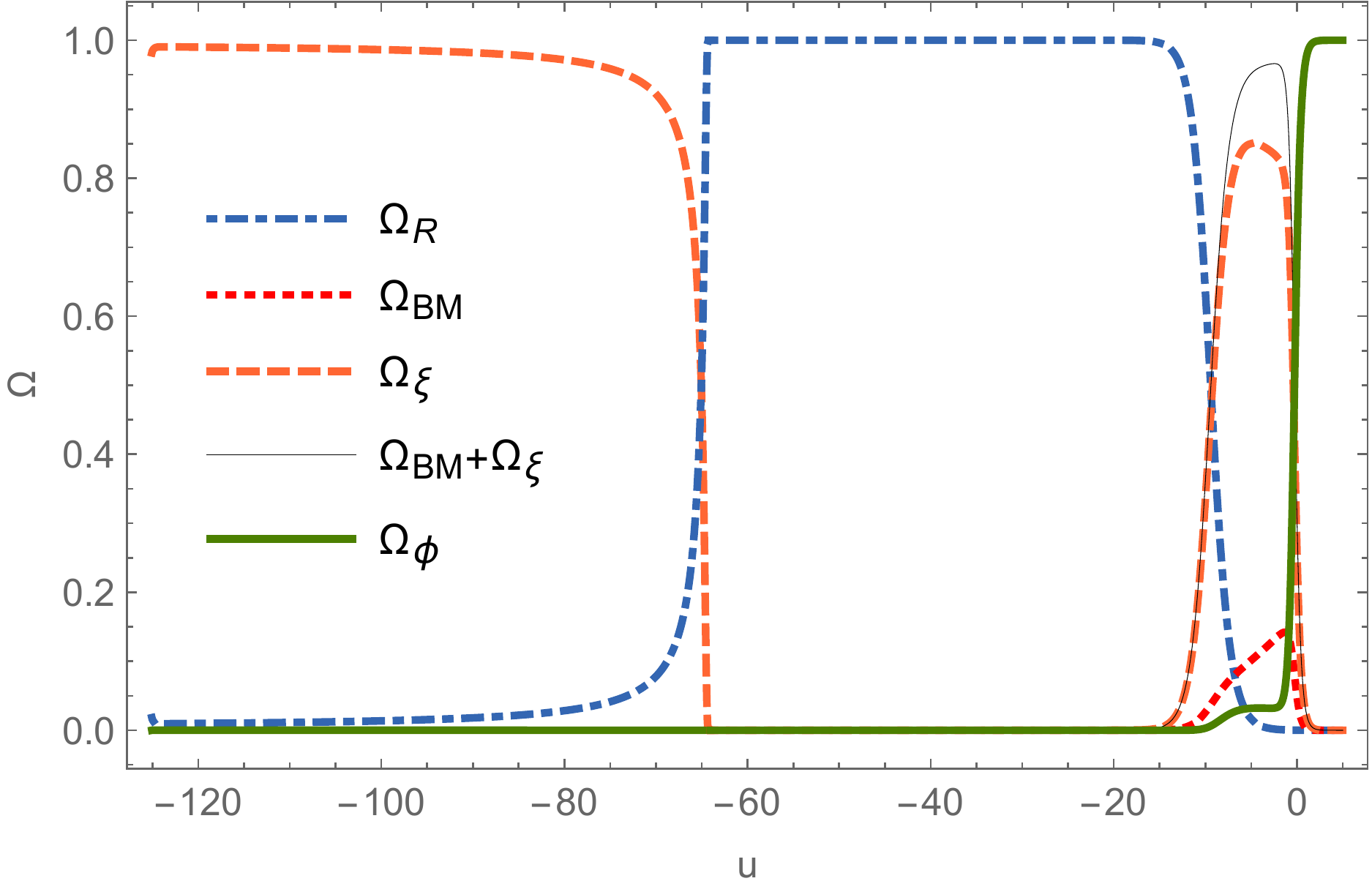}
\caption{Evolution of the density parameters for the case $\alpha=1$ and
$\beta=1/2$. The transition from the radiation- to the matter-dominated era
takes place at $u\approx-9.4$, a little earlier than in the base scenario.
During the matter-dominated era, the density parameter of dark energy is a
non-negligible fraction (about $3\%$)
of the density parameter of matter (dark plus baryonic).
The density parameter of dark matter reaches its maximum value at $u\approx -4.8$,
while for baryonic matter this peaking occurs later, at $u\approx -1.4$.}
 \label{Fig_Omegas-beta05}
\end{figure}

\begin{figure}[t]
\includegraphics[width=7.9cm]{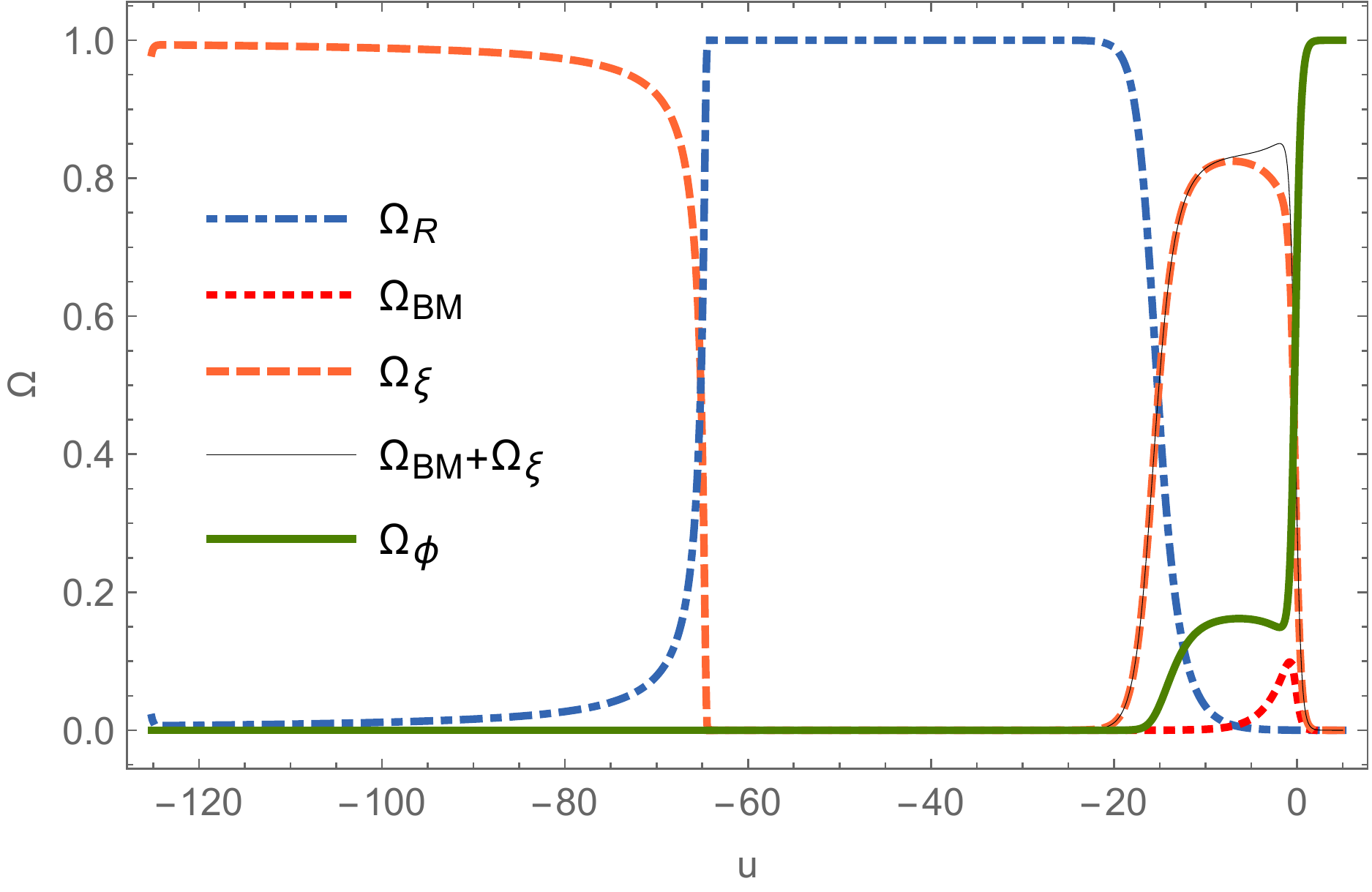}
\caption{Evolution of the density parameters for the case $\alpha=1$ and
$\beta=0$. The transition from the radiation- to the matter-dominated era
takes place at $u\approx-15.2$, much earlier than in the base
scenario.
During the matter-dominated era, the density parameter of dark energy is a
significant fraction (about $20\%$)
of the density parameter of matter (dark plus baryonic).
The density parameter of dark matter reaches its maximum value at $u\approx -7$,
while for baryonic matter this peaking occurs much later, at $u\approx -0.8$.}
 \label{Fig_Omegas-beta0}
\end{figure}

\begin{figure}[t]
\includegraphics[width=7.9cm]{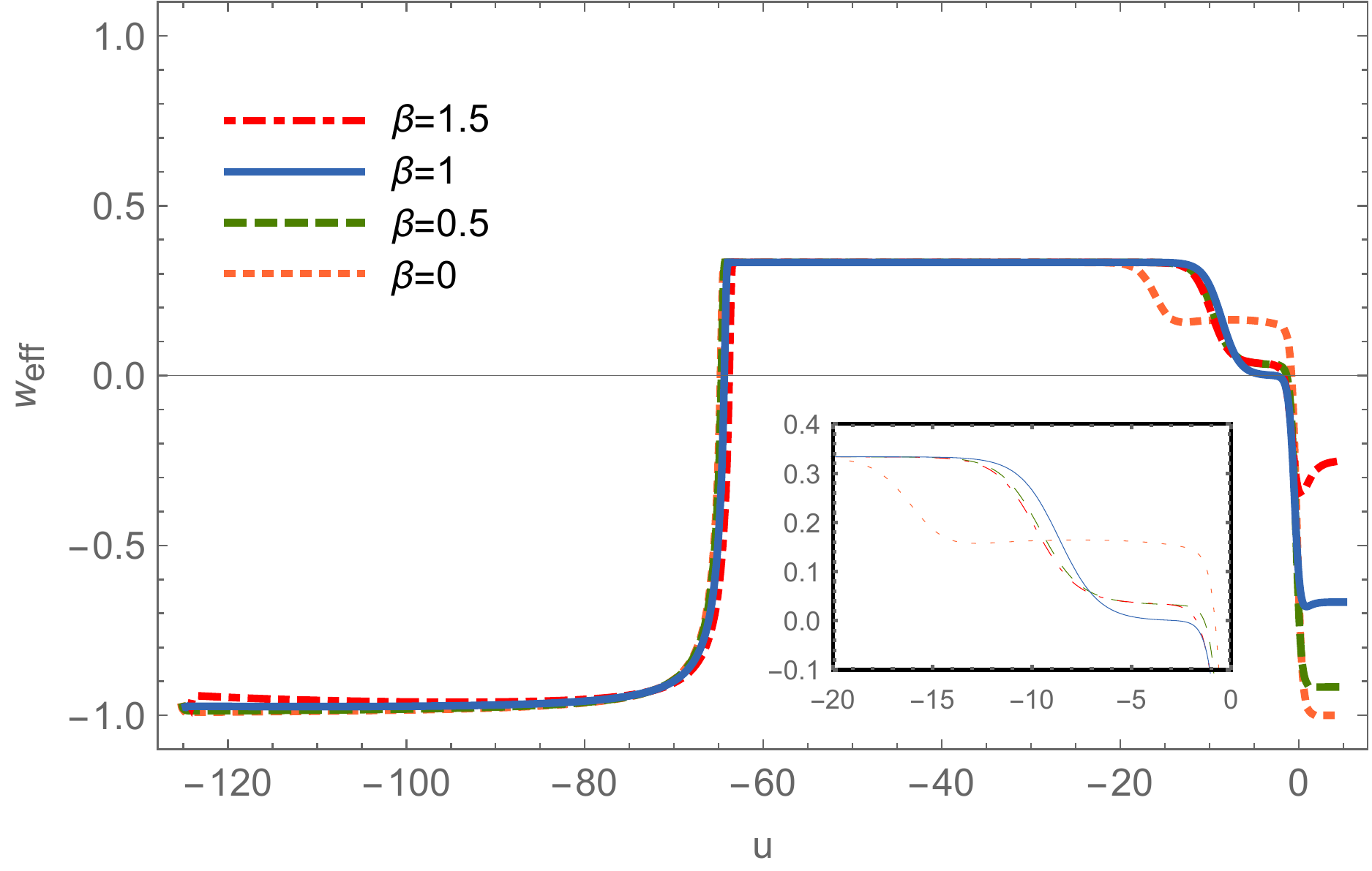}
\caption{Evolution of the effective equation-of-state parameter $w_{\rm eff}$
for $\alpha=1$ and different values of $\beta$ (the solid blue line corresponds
to the base scenario, $\beta=1$).
During the matter-dominated era, the value of $w_{\rm eff}$ is approximately zero
for $\beta=1/2$ and $3/2$, similarly to the base scenario (inset plot).
However, for $\beta=0$, the situation changes drastically, with
the value of $w_{\rm eff}$ departing significantly from zero, due to a
greater influence of dark energy during this era.
For $\beta=0,1/2$, and~$1$, the universe initiates, at recent times, a
period of everlasting accelerated expansion.
For $\beta=3/2$ accelerated expansion takes place at the present time $u_0=0$,
but does not last forever, as $w_{\rm eff}$ asymptotically approaches the value~$-1/4$.}
 \label{Fig-weff-beta}
\end{figure}

Let us start by considering a varying $\beta$ for fixed $\alpha$ (say $\alpha=1$,
as in the base scenario considered in the previous subsection).
The evolution of the density parameters for the scalar fields $\xi$ and $\phi$,
as well as for radiation and baryonic matter, are shown in
Figs.~\ref{Fig_Omegas-beta05} and \ref{Fig_Omegas-beta0} for the cases
$\beta=1/2$ and $\beta=0$, respectively,
while the evolution of the effective equation-of-state parameter
is shown in Fig.~\ref{Fig-weff-beta} for $\beta=0,1/2,1$ and $3/2$.

A first change in the cosmic evolution, as compared with the case $\alpha=\beta$,
is related to the duration of~the matter-dominated era.
The more $\beta$ differs from $\alpha$,
the earlier the transition from a radiation to a matter-dominated universe
takes place and the longer the duration of the latter.
This effect is quite mild for $|\alpha-\beta|\lesssim1$,
having no implications on the viability of the cosmological solutions
(see Fig.~\ref{Fig_Omegas-beta05} and the inset of Fig.~\ref{Fig-weff-beta}).
However, for $|\alpha-\beta|\gtrsim 1$, the effect becomes so strong that it
begins to conflict with primordial nucleosynthesis.
For instance, in the case $\alpha=1$ and $\beta=0$, the transition from the
radiation- to the matter-dominated era takes place already at $u\approx-15.2$
(see Fig.~\ref{Fig_Omegas-beta0}),
quite near to the  primordial nucleosynthesis value $u\approx -18$.
If one further increases the value of $|\alpha-\beta|$,
conflict with primordial nucleosynthesis can only be avoided by dropping the
requirement that the present-time density parameter $\Omega_{\xi}(u_0)$ must be
equal to the observational value $\Omega_\texttt{DM0}$.
In other words, one must accept that the
scalar field $\xi$ accounts only for part of the dark-matter content of the
universe; the rest must be introduced by hand, together with ordinary baryonic
matter. In fact, this was the situation reported in Ref.~\cite{henriques-2009},
where, for $\alpha=0$ and $\beta=-\sqrt{2}$, the scalar field contributed only
about 6\% to the total matter content of the universe.

Another change in the cosmic evolution concerns the behavior of the scalar field
$\phi$ (dark energy).
For $\alpha\neq\beta$, it starts to influence the dynamics of the universe much earlier,
at the beginning of the matter-dominated era, and its energy density is
a non-negligible fraction of the total energy density throughout
the matter-dominated era.
The higher $|\alpha-\beta|$, the greater the fraction of dark energy during this era.
For instance, in the case $\alpha=1$ and $\beta=1/2$,
shown in Fig.~\ref{Fig_Omegas-beta05}, the density parameter of dark energy is about
$3\%$ of the density parameter of matter (dark plus baryonic),
while in the case $\alpha=1$ and $\beta=0$, shown in Fig.~\ref{Fig_Omegas-beta0},
this percentage increases to $20\%$.
This behavior of dark energy implies that, during the matter-dominated era,
the value of the effective equation-of-state parameter $w_{\rm eff}$ differs from zero
(see Fig.~\ref{Fig-weff-beta}).
Note, however, that for the cases $\beta=1/2$ and $3/2$ the dark-energy fraction
of the total energy density during the matter-dominated era is much smaller than
in the case $\beta=0$, meaning that the value of $w_{\rm eff}$
in these cases remains close to zero during the matter-dominated era.

A third change occurring in the cosmic evolution is related to the peaking of
the energy densities of dark and baryonic matter. In the case $\alpha=\beta$,
the energy density of dark energy, given by Eq.~(\ref{rho dark matter}),
evolves exactly as the energy density of ordinary baryonic matter (i.e., as
$e^{-3u}$ or, in terms of the scale factor, as $a^{-3}$), meaning that the
ratio between $\rho_\xi$ and $\rho_\texttt{BM}$ is constant throughout time
and, consequently, the peaking of these two quantities occurs simultaneously.
For $\alpha\neq\beta$ the situation is quite different. The ratio
$\rho_\xi/\rho_\texttt{BM}$ depends directly on the behavior of the scalar
field $\phi$, namely, $\rho_\xi/\rho_\texttt{BM} \propto \exp
[(\alpha-\beta)\kappa\phi/2]$. This implies that the peaking of the energy
densities of dark matter and ordinary baryonic matter does not occur at the
same time. For instance, in the case $\alpha=1$ and $\beta=0$, shown in
Fig.~\ref{Fig_Omegas-beta0}, the density parameter of dark matter reaches its
maximum value at $u\approx -7$, while for baryonic matter this peaking occurs
much later, at $u\approx -0.8$.

As seen above, agreement with current cosmological data requires
$|\alpha-\beta| \lesssim 1$, in order to guarantee that the radiation-dominated era
lasts long enough to encompass primordial nucleosynthesis,
and $|\beta|\lesssim 3/2$, in order to guarantee accelerated expansion at the
present time ($|\beta|<\sqrt{2}$ to guarantee that this accelerated expansion
lasts forever).
For such values of the parameters $\alpha$ and $\beta$,
the two-scalar-field cosmological model given by action~(\ref{action 2SF})
allows for a triple unification of inflation, dark energy, and dark matter
which is, at least qualitatively, consistent with observations.

\begin{figure}[t]
\includegraphics[width=7.9cm]{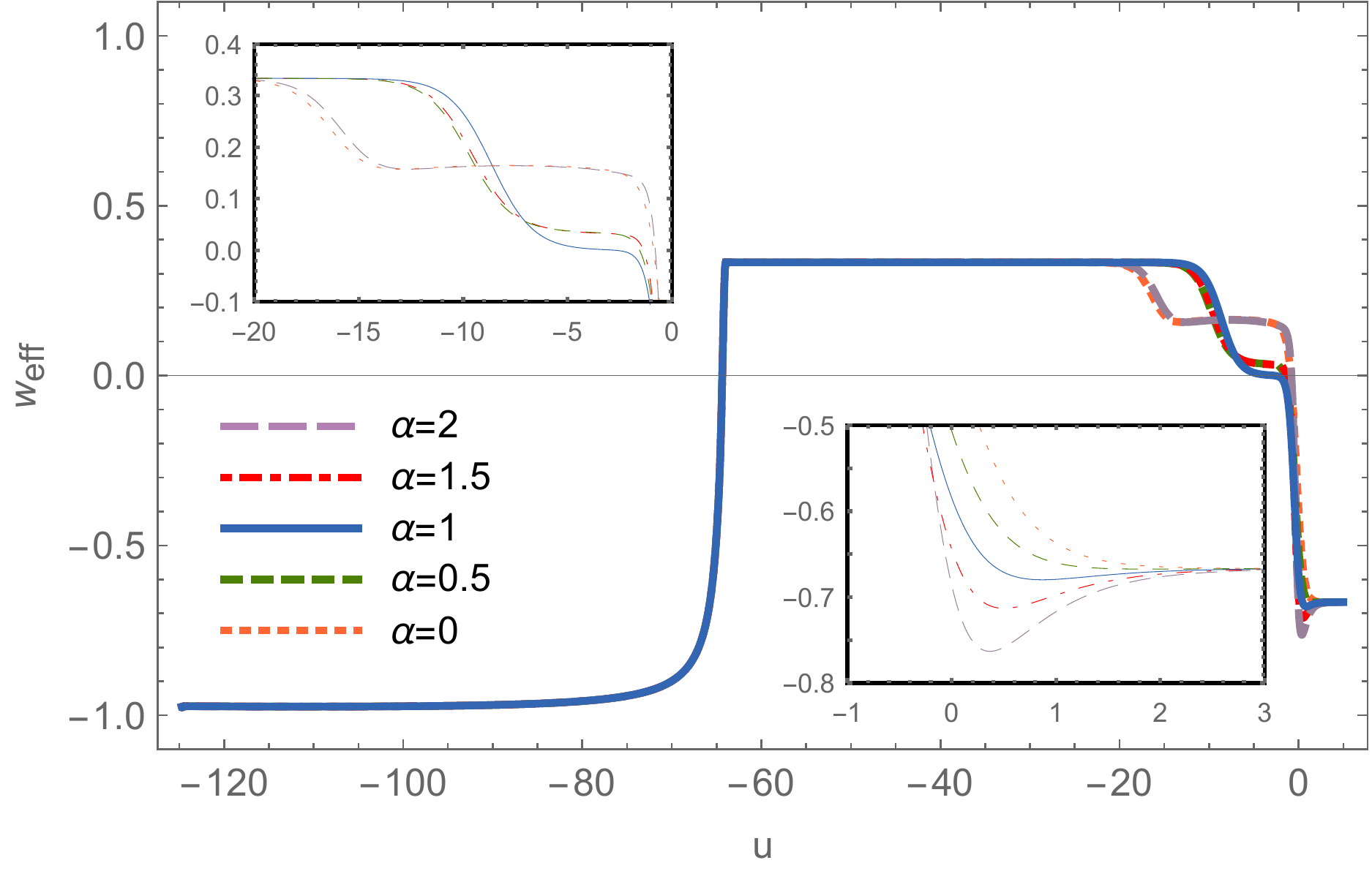}
\caption{Evolution of the effective equation-of-state parameter $w_{\rm eff}$
for $\beta=1$ and different values of $\alpha$ (the solid blue line corresponds
to the base scenario, $\alpha=1$).
At recent times, the universe experiences a second period of accelerated expansion;
as clearly shown in the bottom right inset plot, independently of the value of
parameter $\alpha$, $w_{\rm eff}$ asymptotically approaches $-2/3$.
Due to the earlier influence of dark energy, during the matter-dominated era,
$w_{\rm eff}$ differs significantly from zero in the
cases $\alpha=0$ and $\alpha=2$ and mildly in the cases $\alpha=1/2$ and
$\alpha=3/2$ (see upper left inset plot).}
 \label{Fig-weff-alpha}
\end{figure}

The above conclusions are confirmed by the numerical solutions obtained for
other values of the parameters $\alpha$ and $\beta$.
For completeness, the evolution of the effective equation-of-state parameter for the
case $\beta=1$ and varying $\alpha$ is presented in Fig.~\ref{Fig-weff-alpha}.

\subsection{Dissipative effects \label{num-dissipative}}

We conclude the present section with an analysis of the
dissipative effects during both the inflationary period and the transition
to the radiation-dominated era.

So far we have considered the dissipation coefficients $\Gamma_\xi$ and
$\Gamma_\phi$ to depend linearly on the temperature, i.e., we have chosen
$p=1$ in Eq.~(\ref{gammas}). However, as already referred to in Sect.~\ref{1st
stage}, in the context of warm inflation several other possibilities have been
considered, from the simplest, based on general phenomenological
considerations, to the more elaborate ones, derived from microscopic quantum
field theory. Accordingly, we can extend our analysis to include other values
of $p$, corresponding, in particular, to dissipation coefficients constant
throughout the inflationary period ($p=0$) and dissipation coefficients
inversely proportional to the temperature ($p<0$).

\begin{figure}[t]
\includegraphics[width=7.9cm]{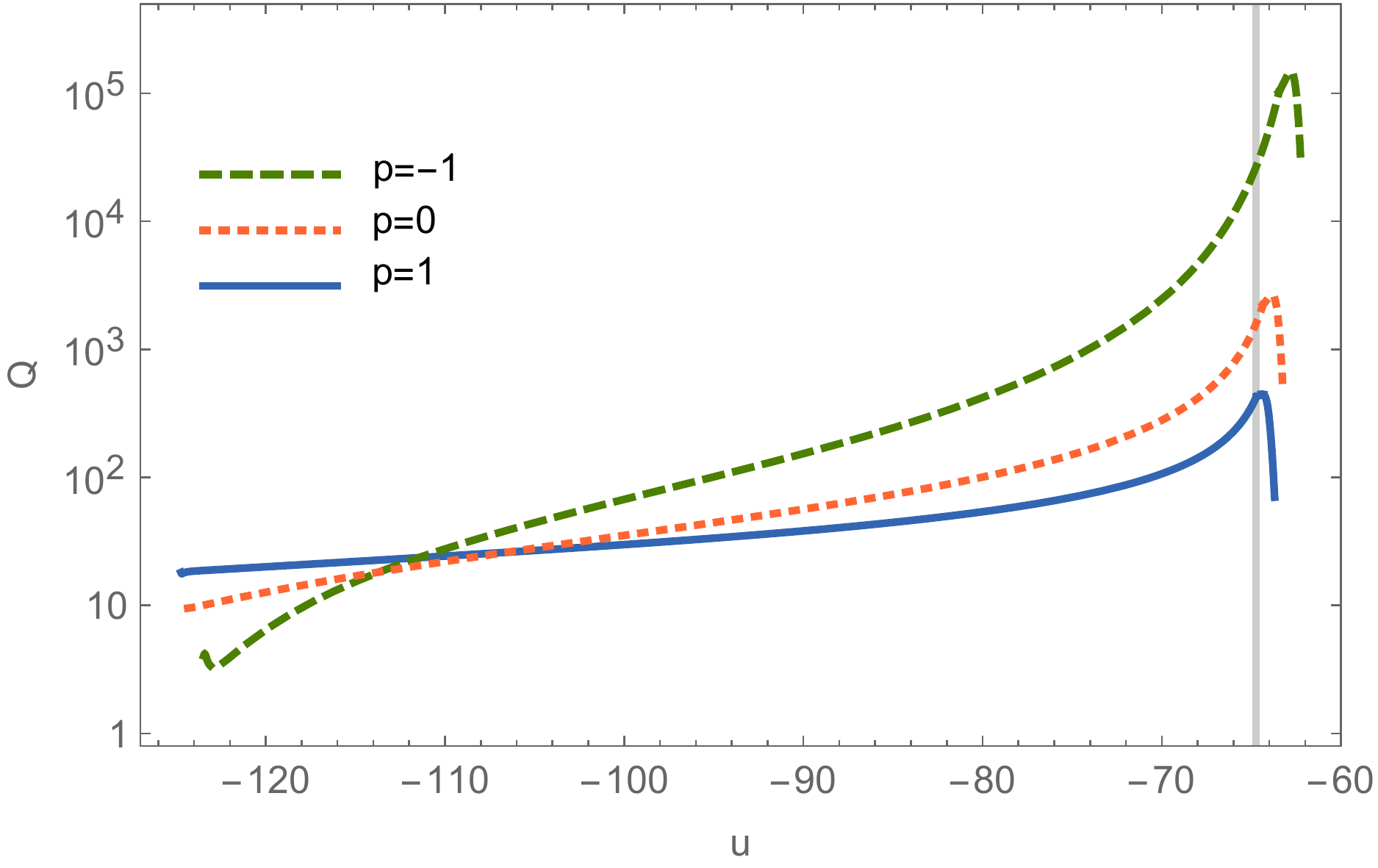}
\caption{Evolution of the dissipation ratio $Q$ for $q=2$ and $p=-1,0,1$,
corresponding to dissipation coefficients inversely proportional to the
temperature, constant, and  proportional to the temperature, respectively.}
 \label{Fig_Gammas_p}
\end{figure}

Our numerical simulations show that, for these values of $p$, the cosmic
evolution proceeds in a similar way to the case $p=1$. In
Fig.~\ref{Fig_Gammas_p} the evolution of the dissipation ratio $Q$ is shown
for $p=-1,0,1$. For comparison purposes, we choose the duration of the
inflationary period to be the same in all three cases, which in turns requires
the choice $f_\xi=f_\phi=2$ for $p=1$, $f_\xi=f_\phi=3.1\times10^{-4}\,
m_\texttt{P}$ for $p=0$, and $f_\xi=f_\phi=3.8\times10^{-8}\, m_\texttt{P}^2$
for $p=-1$ (the initial conditions and the values of the other parameters are
the same as those of the base scenario).

\begin{figure}[t]
\includegraphics[width=7.9cm]{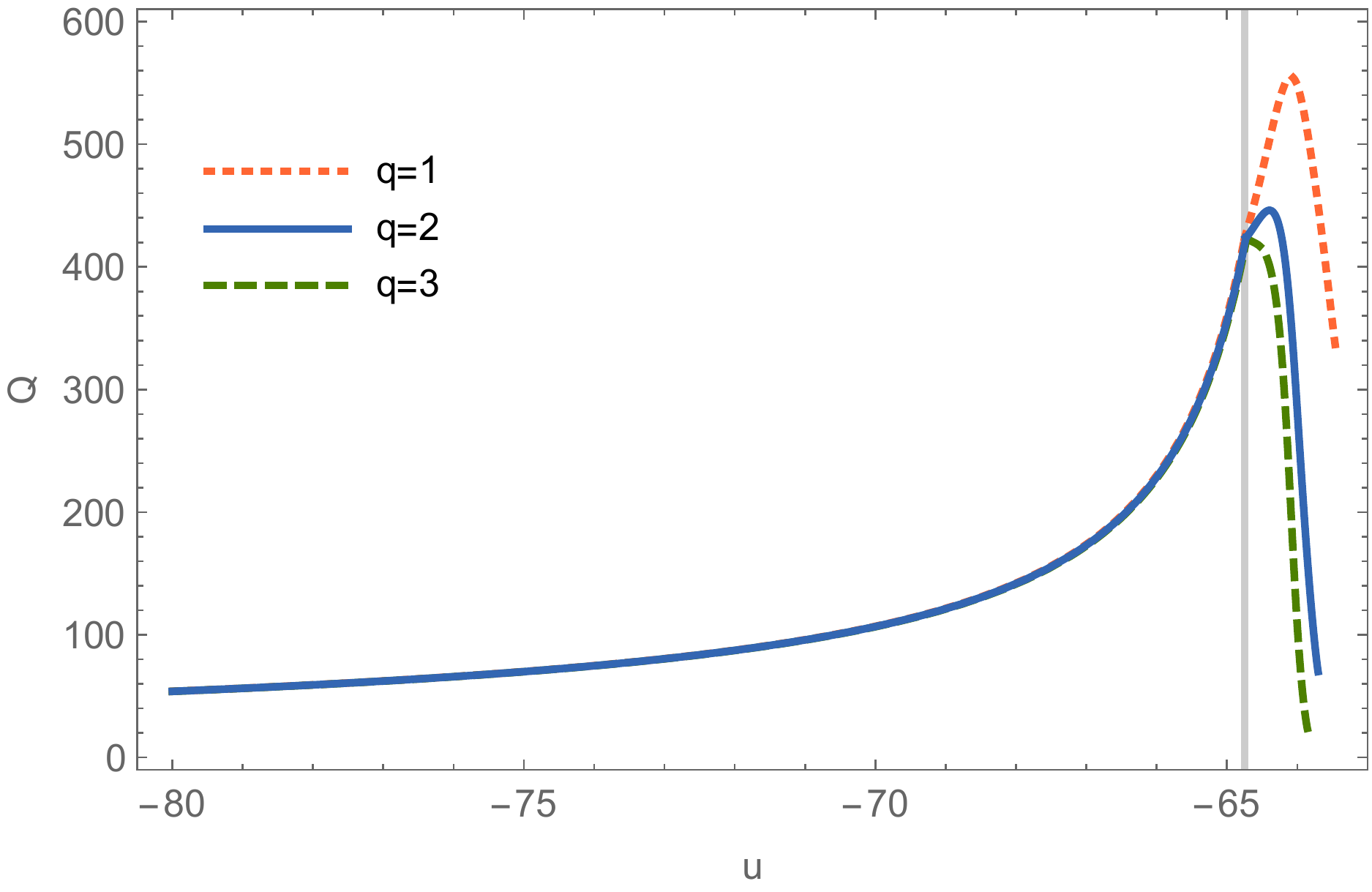}
\caption{Evolution of the dissipation ratio $Q$ for $p=1$ and $q=1,2,3$. The
higher the value of $q$, the faster the dissipation ratio $Q$ is suppressed
after the inflationary period.}
 \label{Fig_Gammas_q}
\end{figure}

In warm inflation, the dissipative effects --- and the consequent energy
transfer from the inflaton to the radiation bath --- play an essential role,
but they should vanish soon after the end of the inflationary period, allowing
cosmic evolution to further proceed in the usual way. In our cosmological
model, this is achieved by assuming that, immediately after the end of
inflation, the dissipation coefficients are suppressed by an exponential term
parameterized by $q$ [see Eq.~(\ref{gammas})]. In all the cases considered so
far, we have choose $q=2$. However, the suppression of dissipative effects
after the inflationary period can be chosen to proceed slower or faster. This
is illustrated in Fig.~\ref{Fig_Gammas_q}, where the dissipation ratio $Q$ is
shown for different values of $q$, namely $q=1,2,3$, with $p=1$ and
$f_\xi=f_\phi=2$.

Let us note that for $p>2$, the dissipation coefficients $\Gamma_\xi$ and
$\Gamma_\phi$ are suppressed naturally after inflation, with no need for an
explicit exponential suppression term in Eq.~(\ref{gammas}). This can be seen
as follows.

\begin{figure}[t]
\includegraphics[width=7.9cm]{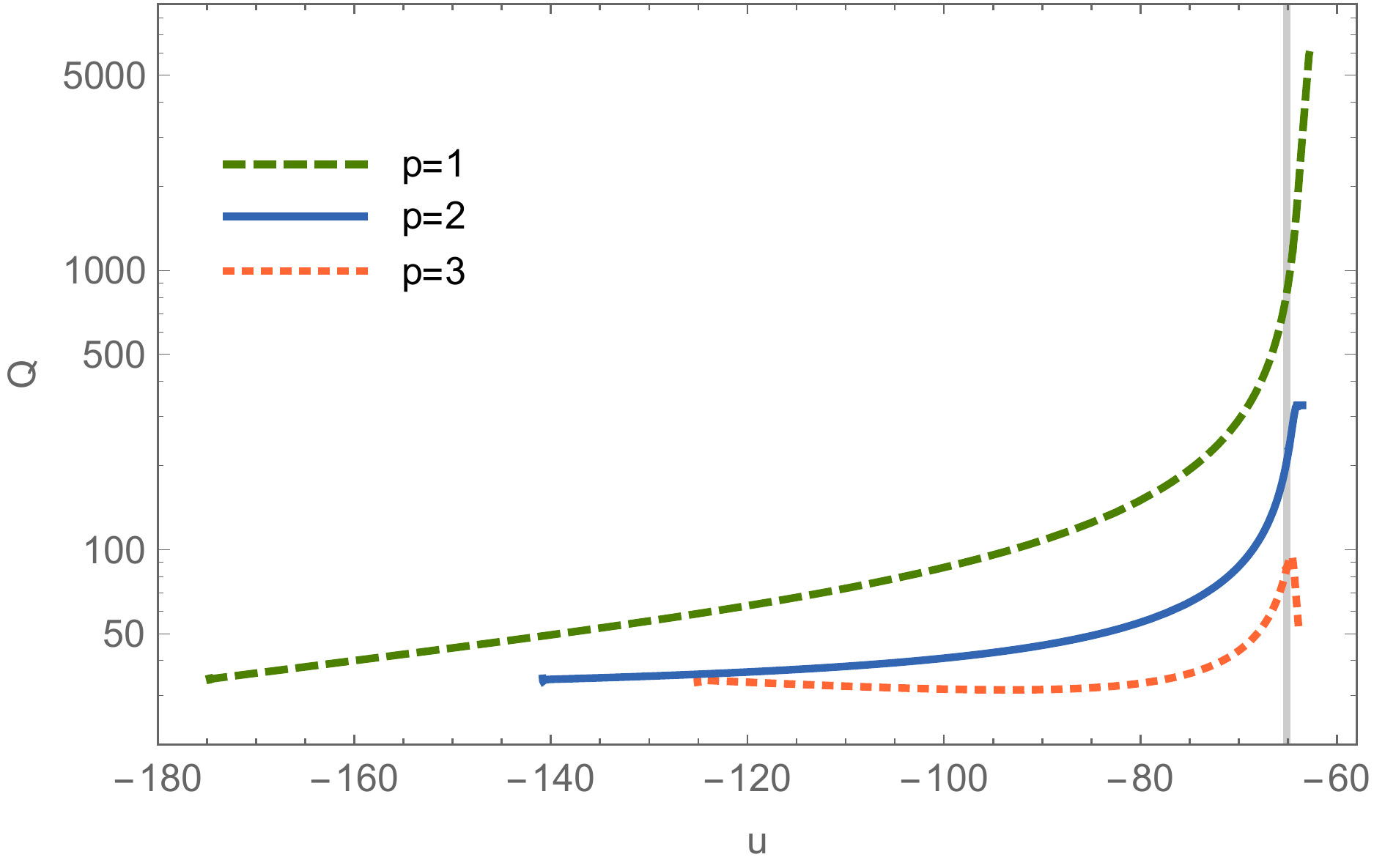}
\caption{Evolution of the dissipation ratio $Q$ for $q=0$ and $p=1,2,3$. For
$p>2$, as a result of the background dynamics, suppression takes place even in
the absence of the exponential factor in Eq.~(\ref{gammas}).}
 \label{Fig_Gammas_q0}
\end{figure}

After the end of the inflationary period ($u>u_\texttt{E}$), cosmic evolution is
dominated by radiation.
Let $\tilde{u}>u_\texttt{E}$ be the value of $u$ above which the dissipative terms in
Eq.~(\ref{Eq rho E1}) are much smaller than the energy density of radiation,
i.e., $(\dot{a}/a)(\Gamma_\xi \xi_u^2 + \Gamma_\phi \phi_u^2) \ll
\rho_\texttt{R}$. Then, for $u>\tilde{u}$, the energy density of radiation
evolves approximately as
$\rho_\texttt{R}(u)\simeq\tilde{\rho}_\texttt{R}e^{-4(u-\tilde{u})}$, where
$\tilde{\rho}_\texttt{R} \equiv \rho_\texttt{R}(\tilde{u})$. Inspection of the
numerical solutions of Eq.~(\ref{Eq friedman E1}) reveals that, for
$u>\tilde{u}$, the Hubble parameter is given, in a good approximation, by $H
\simeq \kappa (\rho_\texttt{R}/3)^{1/2}$. Then, from Eqs.~(\ref{gammas})
and (\ref{q}) one obtains an approximate analytical
expression for the dissipation ratio as a function of $u$, valid for
$u>\tilde{u}$,
\begin{align}
 Q \simeq
 A_1 \exp \left[ 1-(p-2)(u-\tilde{u})- A_2 e^{q(u-\tilde{u})} \right],
   \label{Q-approx}
\end{align}
where $A_1$ and $A_2$ are given by
\begin{align}
 A_1 = {}& \frac{f_{\xi,\phi}}{\sqrt3\kappa}
  \left( \frac{30}{\pi^2 g_*} \right)^{\frac{p}{4}}
  \tilde{\rho}_\texttt{R}^{\frac{p-2}{4}}, \\
 A_2 = {}& \left( \frac{\rho_\texttt{RE}}{\tilde{\rho}_\texttt{R}}
           \right)^{\frac{q}{4}},
\end{align}
and $\rho_\texttt{RE} \equiv \rho_\texttt{R}(u_\texttt{E})$ denotes the energy
density of radiation at the end of the inflationary period. For $p\leq2$, the
dissipation ratio $Q$ does not decrease if $q=0$, but for $p>2$ the parameter
$q$ can be set to zero and suppression nevertheless takes place (see
Fig.~\ref{Fig_Gammas_q0}). This result is in agreement with
Ref.~\cite{lima-2019}, where, in the context of a quintessential inflationary
model, the dissipation coefficient, assumed to depend both on the temperature
and the inflaton field as $\Gamma\propto T^p \xi^c$, is shown to be a
decreasing function of the number of $e$-folds for $p>2$.

So far, the cases considered correspond to a strong dissipative regime, for which
$Q>1$. However, it is of relevance to show that a weak dissipative regime can also
be obtained in our two-scalar-field cosmological model for other choices of the
initial conditions and the values of the parameters.
Indeed, as already mentioned above, the confrontation of warm-inflation predictions
with the latest CMB data
\cite{bastero-gil-2019, benetti-2017,arya-2018,bastero-gil-2018,bastero-gil-2018a,motaharfar-2019}
has shown a preference for the weak dissipation regime for certain combinations
of the inflaton's potential and the dissipation coefficients (for instance,
for a quartic potential and a dissipation coefficient proportional to $T$ or $T^3$).

\begin{figure}[t]
\includegraphics[width=7.9cm]{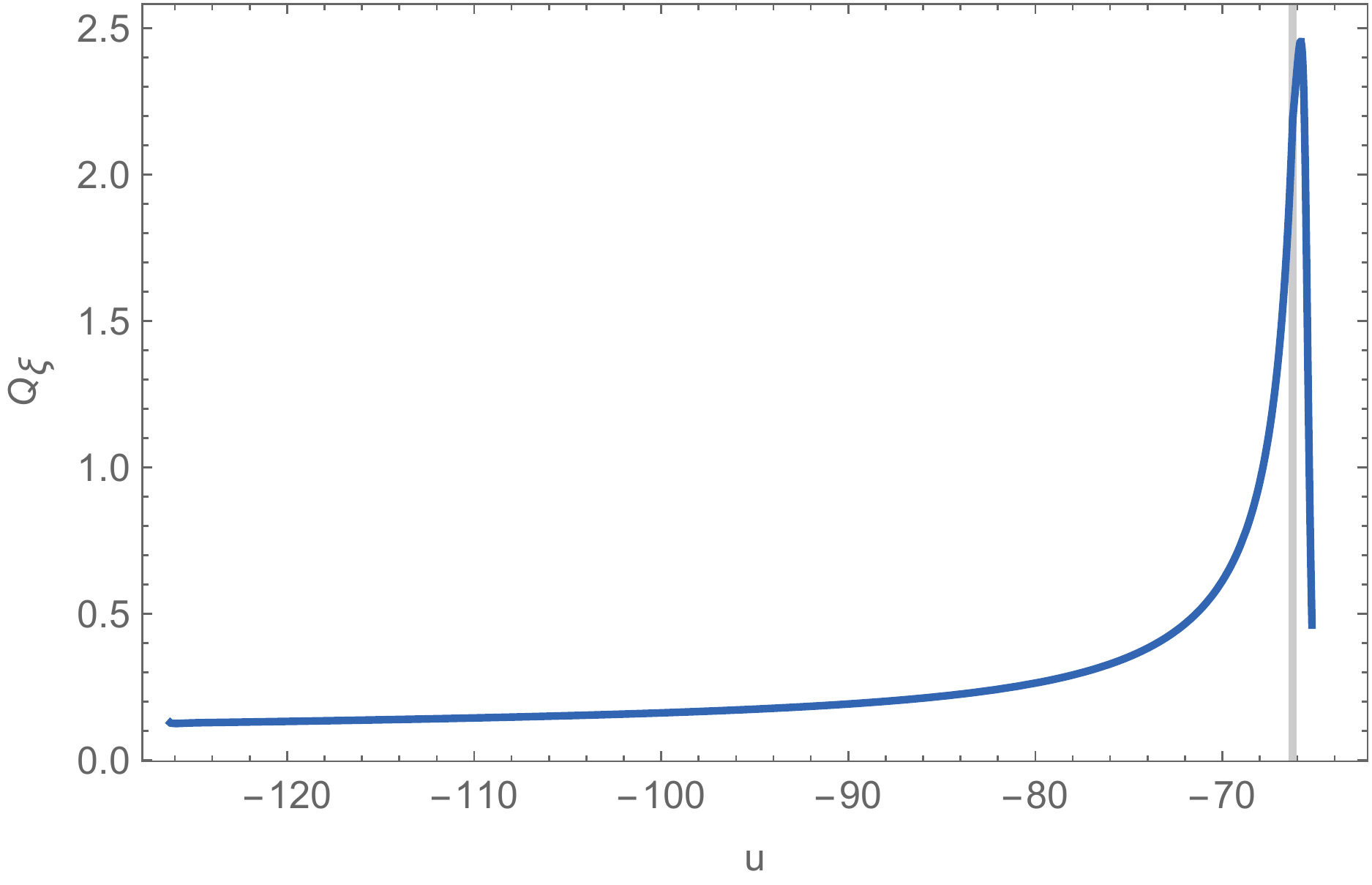}
\caption{Evolution of the dissipation ratio $Q_\xi$. Weak dissipation is maintained
throughout most of the inflationary period. Towards its end, the dissipation ratio
sharply increases, reaching values above unity, which facilitates the transition to
the radiation-dominated era.}
 \label{Fig_Q_xi}
\end{figure}

From Eq.~(\ref{q}), it follows that a smaller value of the dissipation ratio
at the onset of the first stage of evolution, $Q_\xi(u=u_i)$, can be obtained
by decreasing $f_\xi$ and/or increasing $\xi(u_i)$ and $\rho_\texttt{R}(u_i)$.
Starting from the base scenario, considered in detail in Sect.~\ref{num-base},
it is then straightforward to find a set of initial conditions and values of
the parameters corresponding to weak dissipation, as, for instance,
$\xi(u_i)=3.19\, m_\texttt{P}$, $\phi(u_i)=10^{-3}\, m_\texttt{P}$,
$\xi_u(u_i)=10^{-2}\, m_\texttt{P}$, $\phi_u(u_i)=10^{-5}\, m_\texttt{P}$,
$\rho_\texttt{R}(u_i)=2.2 \times 10^{-12}\, m_\texttt{P}^4$, $V_a=2.69 \times
10^{-123}\, m_\texttt{P}^4$, $m=10^{-5}\, m_\texttt{P}$, $f_\xi=0.05$,
$f_\phi=25$, $p=1$, and $q=2$. The evolution of the dissipation ratio $Q_\xi$
for this case is shown in Fig.~\ref{Fig_Q_xi} (the effective equation-of-state
parameter and the density parameters evolve similarly as in the base scenario,
making it unnecessary to present them here).

The primordial spectrum of density perturbations of our unification model and
its agreement with observational data, for the quadratic potential, given by
Eq.~(\ref{potential xi}), and the dissipation coefficients, given by
Eq.~(\ref{gammas}), both in the strong and weak dissipative regimes,
will be explored in future work.

\section{Conclusions\label{conclusions}}

In this article we have presented a unified description of inflation, dark energy,
and dark matter in a two-scalar-field cosmological model with a non-standard
kinetic term and an exponential potential.
Such models arise in a great variety of gravity theories, such as the
Jordan-Brans-Dicke theory, Kaluza-Klein theories, $f(R)$-gravity, string theories,
and hybrid metric-Palatini theories of gravity.

In the proposed triple unification, one of the scalar fields plays the role of
inflaton and dark matter and the other plays the role of dark energy.
More specifically, inflation, assumed to be of the warm type, is driven by the
scalar field $\xi$, which, shortly after the end of the inflationary period,
decouples from radiation and begins to oscillate rapidly around the minimum
of its potential, thus behaving like a cold-dark-matter fluid; the second scalar
field $\phi$ emerges, at recent times, as the dominant component of the universe
and gives rise to an era of accelerated expansion. Thus, seemingly disparate
phenomena like inflation, dark energy, and dark matter are unified under the
same theoretical roof using scalar fields.

The two-scalar-field cosmological model given by action~(\ref{action 2SF})
contains two dimensionless parameters, $\alpha$ and $\beta$ (for $\alpha=0$,
the kinetic term for the scalar field $\xi$ becomes canonical; for $\beta=0$,
the direct coupling in the potential between the two scalar fields $\xi$
and $\phi$ disappears). These parameters could, in principle, be chosen freely;
however, as detailed in Sect.~\ref{num-deviations}, the requirement that the
transition from the radiation- to the matter-dominated era does not occur too
early in the cosmic history and, consequently, does not conflict with primordial
nucleosynthesis, as well as the requirement that the expansion of the universe is
accelerating at the present time, imposes constraints on the parameters $\alpha$
and $\beta$, namely, $|\alpha-\beta| \lesssim 1$
and $|\beta| \lesssim 3/2$.

For such values of $\alpha$ and $\beta$, the picture emerging in our unified
description of inflation, dark energy, and dark matter is consistent with the
standard cosmological model.
Indeed, the inflationary period is followed by a radiation-dominated era that
encompasses the primordial nucleosynthesis epoch; the matter-dominated era lasts
long enough for structure formation to occur; the transition to a
dark-energy-dominated universe takes place in a recent past;
and the density parameters for dark matter and dark energy, as well as for radiation
and ordinary baryonic matter, evaluated at the present time, are in agreement with
current cosmological observations (see Fig.~\ref{Fig-Omega} for the case $\alpha=\beta=1$,
Fig.~\ref{Fig_Omegas-beta05} for the case $\alpha=1$, $\beta=1/2$,
and Fig.~\ref{Fig_Omegas-beta0} for the case $\alpha=1$, $\beta=0$).

Of crucial importance for consistency with the standard cosmological model is
the value of the energy density of the scalar field $\xi$ at the moment when
this field begins to oscillate rapidly around its minimum, changing its
behavior from an inflaton field to a nonrelativistic dark-matter fluid (i.e.,
at the transition from the first to second stage of cosmic evolution). If this
energy density is too large, the radiation-dominated era is too short,
conflicting with primordial nucleosynthesis; if it is too small, the
matter-dominated era is not long enough for structure formation to take place
(see Fig.~\ref{Fig_rhoxi-rhoR-E1}). In our model, an appropriate value of the
energy density of the inflaton/dark-matter field at the transition between the
first and second stages of evolution is achieved by ensuring that, immediately
after the end of the inflationary period, the dissipation coefficients,
responsible for the energy transfer from the scalar fields $\xi$ and $\phi$ to
the radiation bath, are rapidly suppressed, becoming negligible soon
afterwards. When this happens, the scalar field $\xi$ decouples from radiation
and, because its effective mass is larger than the Hubble parameter, begins to
oscillate rapidly around its minimum, behaving like dark matter.

In our cosmological model, the behavior of the dissipation coefficients is
controlled by two parameters, $p$ and $q$. The first sets the temperature
dependence of the coefficients and the second determines how fast these
coefficients are suppressed after the end of the inflationary period [see
Eq.~(\ref{gammas})]. As we have shown, for $p>2$, the dissipation coefficients
are suppressed naturally after inflation, with no need for an explicit
suppression term, allowing us to set $q=0$. This does not mean, however, that
the models with $p\leq2$, requiring $q\neq0$, are less admissible. Actually,
such models, have been proposed in recent years
\cite{bastero-gil-2016,rosa-2019,rosa-2019b,bastero-gil-2019} and are quite
sound, both theoretically and observationally.

As already mentioned above, shortly after the end of the inflationary period,
the scalar field $\xi$ begins to oscillate rapidly around the minimum of its
potential. We have derived an expression for its energy density during this
oscillating phase [see Eq.~(\ref{rho dark matter})],
showing that it is proportional to
$\exp[(\alpha-\beta)\kappa\phi/2]$.
In the case $\alpha=\beta$, this exponential equals unity and dark matter behaves
exactly as ordinary baryonic matter, i.e., evolves as $a^{-3}$, where $a$ is
the scale factor. But for $\alpha\neq\beta$ the situation is quite different;
the energy density of dark matter depends directly on the dark-energy field $\phi$,
leading to a non-simultaneous peaking of the energy densities of dark matter and
ordinary baryonic matter
(see Figs.~\ref{Fig_Omegas-beta05} and \ref{Fig_Omegas-beta0}).
Furthermore, in the case $\alpha\neq\beta$, the energy density of the dark energy
is a non-negligible fraction of the critical energy density throughout the
matter-dominated era.

Finally, we have shown that the effective equation-of-state parameter $w_{\rm
eff}$ tends, asymptotically, to $-1+\beta^2/3$, implying that, for
$|\beta|<\sqrt2$, the universe enters a period of everlasting accelerated
expansion (for values of $|\beta|$ slightly above $\sqrt2$, this accelerated
expansion still takes place, but does not last forever).

In this article, we have proposed a triple unification of inflation, dark
energy, and dark matter in a two-scalar-field cosmological model. This is a
first approach, intended to show that such a unification is, in principle,
possible and that it reproduces, at least qualitatively, the main features of
the observed universe. We expect to explore and deepen this model in future
publications.


\begin{thebibliography}{99}

\bibitem{starobinsky-1980}
A.~A.~Starobinsky,
\textit{A new type of isotropic cosmological models without singularity},
Phys.\ Lett.\ B \textbf{91}, 99--102 (1980).

\bibitem{guth-1981}
A.~H.~Guth,
\textit{Inflationary universe: A possible solution to the horizon and flatness
problems},
Phys.\ Rev.\ D \textbf{23}, 347--356 (1981).

\bibitem{linde-1982}
A.~D.~Linde,
\textit{A new inflationary universe scenario: A possible solution of the horizon,
flatness, homogeneity, isotropy and primordial monopole problems},
Phys.\ Lett.\ B \textbf{108}, 389--393 (1982).

\bibitem{albrecht-1982}
A.~Albrecht and P.~J.~Steinhardt,
\textit{Cosmology for grand unified theories with radiatively induced symmetry breaking},
Phys.\ Rev.\ Lett.\ \textbf{48}, 1220--1223 (1982).

\bibitem{linde-1983}
A.~D.~Linde, \textit{Chaotic inflation}, Phys.\ Lett.\ B \textbf{129}, 177-181
(1983).

\bibitem{Planck-inflation-2015}
P.~A.~R.~Ade {\it et al.} (Planck Collaboration),
\textit{Planck 2015 results: XX.\ Constraints on inflation},
Astron.\ Astrophys.\ \textbf{594}, A20 (2016).

\bibitem{riess-1998}
A.~G.~Riess {\it et al.} (Supernova Search Team),
\textit{Observational evidence from supernovae for an accelerating universe and
a cosmological constant},
Astron.\ J.\ \textbf{116}, 1009--1038 (1998).

\bibitem{perlmutter-1999}
S.~Perlmutter {\it et al.} (The Supernova Cosmology Project),
\textit{Measurements of $\Omega$ and $\Lambda$ from 42 high-redshift supernovae},
Astrophys.\ J.\ \textbf{517}, 565--586 (1999).

\bibitem{Planck-parameters-2015}
P.~A.~R.~Ade {\it et al.} (Planck Collaboration),
\textit{Planck 2015 results: XIII.\ Cosmological parameters},
Astron.\ Astrophys.\ \textbf{594}, A13 (2016).

\bibitem{weinberg-1989}
S.~Weinberg,
\textit{The cosmological constant problem},
Rev.\ Mod.\ Phys.\ \textbf{61}, 1--23 (1989).

\bibitem{caldwell-1998}
R.~R.~Caldwell, R.~Dave, and P.~J.~Steinhardt,
\textit{Cosmological imprint of an energy component with general equation of state},
Phys.\ Rev.\ Lett.\ \textbf{80}, 1582--1585 (1998).

\bibitem{bertone-2018}
G.~Bertone and T.~M.~P.~Tait,
\textit{A new era in the search for dark matter},
Nature \textbf{562}, 51--56 (2018).

\bibitem{liddle-2006}
A.~R.~Liddle and L.~A.~Ure\~na-L{\'o}pez,
\textit{Inflation, dark matter, and dark energy in the string landscape},
Phys.\ Rev.\ Lett.\ \textbf{97}, 161301 (2006).

\bibitem{liddle-2008}
A.~R.~Liddle, C.~Pahud, and L.~A.~Ure\~na-L{\'o}pez,
\textit{Triple unification of inflation, dark matter, and dark energy using a single field},
Phys.\ Rev.\ D \textbf{77}, 121301(R) (2008).

\bibitem{henriques-2009}
A.~B.~Henriques, R.~Potting, and P.~M.~S{\'a},
\textit{Unification of inflation, dark energy, and dark matter within the
Salam-Sezgin cosmological model},
Phys.\ Rev.\ D \textbf{79}, 103522 (2009).

\bibitem{berera-1995}
A.~Berera,
\textit{Warm inflation},
Phys.\ Rev.\ Lett.\ \textbf{75}, 3218--3221 (1995).

\bibitem{berkin-1991}
A.~L.~Berkin and K.~I.~Maeda,
\textit{Inflation in generalized Einstein theories},
Phys.\ Rev.\ D \textbf{44}, 1691--1704 (1991).

\bibitem{starobinsky-2001}
A.~A.~Starobinsky, S.~Tsujikawa, and J.~Yokoyama,
\textit{Cosmological perturbations from multi-field inflation in generalized Einstein theories},
Nucl.\ Phys.\ B \textbf{610}, 383--410 (2001).

\bibitem{harko-2012}
T.~Harko, T.~S.~Koivisto, F.~S.~N.~Lobo, and G.~J.~Olmo,
\textit{Metric-Palatini gravity unifying local constraints and late-time cosmic acceleration},
Phys.\ Rev.\ D \textbf{85}, 084016 (2012).

\bibitem{tamanini-2013}
N.~Tamanini and C.~G.~B\"{o}hmer,
\textit{Generalized hybrid metric-Palatini gravity},
Phys.\ Rev.\ D \textbf{87}, 084031 (2013).

\bibitem{capozziello-2006}
S.~Capozziello, S.~Nojiri, and S.~D.~Odintsov,
\textit{Unified phantom cosmology: Inflation, dark energy and dark matter under
the same standard},
Phys.\ Lett.\ B \textbf{632}, 597–604 (2006).

\bibitem{bose-2009}
N.~Bose and A.~S.~Majumdar,
\textit{k-essence model of inflation, dark matter, and dark energy},
Phys.\ Rev.\ D \textbf{79}, 103517 (2009).

\bibitem{santiago-2011}
J.~De-Santiago and J.~L.~Cervantes-Cota,
\textit{Generalizing a unified model of dark matter, dark energy, and inflation
with a noncanonical kinetic term},
Phys.\ Rev.\ D \textbf{83}, 063502 (2011).

\bibitem{odintsov-2019}
S.~D.~Odintsov and V.~K.~Oikonomou,
\textit{Unification of inflation with dark energy in $f(R)$ gravity and
axion dark matter},
Phys.\ Rev.\ D \textbf{99}, 104070 (2019).

\bibitem{lima-2019}
G.~B.~F.~Lima and R.~O.~Ramos,
\textit{Unified early and late universe cosmology through dissipative effects in steep quintessential inflation potential models},
Phys.\ Rev.\ D \textbf{100}, 123529 (2019).

\bibitem{ketov-2020}
S.~V.~Ketov,
\textit{Supergravity as the dark side of the universe},
Int.\ J.\ Mod.\ Phys.\ A \textbf{35}, 2040038 (2020).

\bibitem{odintsov-2020}
S.~D.~Odintsov and V.~K.~Oikonomou, \textit{Geometric inflation and dark
energy with axion $F(R)$ gravity}, Phys.\ Rev.\ D \textbf{101}, 044009 (2020).

\bibitem{berera-2009}
A.~Berera, I.~G.~Moss, and R.~O.~Ramos,
\textit{Warm inflation and its microphysical basis},
Rept.\ Prog.\ Phys.\ \textbf{72}, 026901 (2009).

\bibitem{bastero-gil-2009}
M.~Bastero-Gil and A.~Berera,
\textit{Warm inflation model building},
Int.\ J.\ Mod.\ Phys.\ A \textbf{24}, 2207--2240 (2009).

\bibitem{bastero-gil-2013}
M.~Bastero-Gil, A.~Berera, R.~O.~Ramos, and J.~G.~Rosa,
\textit{General dissipation coefficient in low-temperature warm inflation},
J.\ Cosmol.\ Astropart.\ Phys.\ \textbf{01}, 016 (2013).

\bibitem{bastero-gil-2016}
M.~Bastero-Gil, A.~Berera, R.~O.~Ramos, and J.~G.~Rosa,
\textit{Warm little inflaton},
Phys.\ Rev.\ Lett.\ \textbf{117}, 151301 (2016).

\bibitem{rosa-2019}
J.~G.~Rosa and L.~B.~Ventura,
\textit{Warm little inflaton becomes cold dark matter},
Phys.\ Rev.\ Lett.\ \textbf{122}, 161301 (2019).

\bibitem{rosa-2019b}
J.~G.~Rosa and L.~B.~Ventura,
\textit{Warm little inflaton becomes dark energy},
Phys.\ Lett.\ B \textbf{798}, 134984 (2019).

\bibitem{bastero-gil-2019}
M.~Bastero-Gil, A.~Berera, R.~O.~Ramos, and J.~G.~Rosa,
\textit{Towards a reliable effective field theory of inflation},
arXiv:1907.13410 [hep-th].

\bibitem{benetti-2017}
M.~Benetti and R.~O.~Ramos,
\textit{Warm inflation dissipative effects: Predictions and constraints
from the Planck data},
Phys.\ Rev.\ D \textbf{95}, 023517 (2017).

\bibitem{arya-2018}
R.~Arya, A.~Dasgupta, G.~Goswami, J.~Prasad, and R.~Rangarajan,
\textit{Revisiting CMB constraints on warm inflation},
J.\ Cosmol.\ Astropart.\ Phys.\ \textbf{02}, 043 (2018).

\bibitem{bastero-gil-2018}
M.~Bastero-Gil, S.~Bhattacharya, K.~Dutta, and M.~R.~Gangopadhyay,
\textit{Constraining warm inflation with CMB data},
J.\ Cosmol.\ Astropart.\ Phys.\ \textbf{02}, 054 (2018).

\bibitem{bastero-gil-2018a}
M.~Bastero-Gil, A.~Berera, R.~Hern{\'a}ndez-Jim{\'e}nez, and J.~G.~Rosa,
\textit{Dynamical and observational constraints on the warm little inflaton scenario},
Phys.\ Rev.\ D \textbf{98}, 083502 (2018).

\bibitem{motaharfar-2019}
M.~Motaharfar, V.~Kamali, and R.~O.~Ramos,
\textit{Warm inflation as a way out of the swampland},
Phys.\ Rev.\ D \textbf{99}, 063513 (2019).

\bibitem{turner-1983}
M.~S.~Turner,
{\it Coherent scalar-field oscillations in an expanding universe},
Phys.\ Rev.\ D \textbf{28}, 1243--1247 (1983).

\bibitem{sa-2020}
P.~M.~S{\'a},
\textit{Unified description of dark energy and dark matter within the
generalized hybrid metric-Palatini theory of gravity},
Universe \textbf{6}, 78 (2020).

\end{thebibliography}
\end{document}